\documentclass[11pt, a4paper]{article}
\usepackage{jcappub, bm}

\newcommand{\nhat}{\mathbf{\hat{n}}}

\title{Testing $\bm{\Lambda}$CDM at the lowest redshifts with SN Ia and galaxy velocities}

\author[a]{Dragan Huterer,}
\author[a,b]{Daniel L. Shafer,}
\author[c,d]{Daniel M. Scolnic}
\author[e]{and Fabian~Schmidt}

\affiliation[a]{Department of Physics, University of Michigan,\\
450 Church Street, Ann Arbor, MI 48109, U.S.A.}
\affiliation[b]{Department of Physics and Astronomy, Johns Hopkins University,\\
3400 North Charles Street, Baltimore, MD 21218, U.S.A.}
\affiliation[c]{University of Chicago, Kavli Institute for Cosmological Physics,\\
5640 South Ellis Avenue, Chicago, IL 60637, U.S.A.}
\affiliation[d]{Hubble, KICP fellow}
\affiliation[e]{Max-Planck-Institut f\"ur Astrophysik,\\
Karl-Schwarzschild-Str.\ 1, 85748 Garching, Germany}

\emailAdd{huterer@umich.edu}
\emailAdd{dshafer2@jhu.edu}
\emailAdd{dscolnic@kicp.uchicago.edu}
\emailAdd{fabians@mpa-garching.mpg.de}

\abstract{Peculiar velocities of objects in the nearby universe are correlated due to the gravitational pull of large-scale structure. By measuring these velocities, we have a unique opportunity to test the cosmological model at the lowest redshifts. We perform this test, using current data to constrain the amplitude of the ``signal'' covariance matrix describing the velocities and their correlations. We consider a new, well-calibrated ``Supercal'' set of low-redshift SNe Ia as well as a set of distances derived from the fundamental plane relation of 6dFGS galaxies. Analyzing the SN and galaxy data separately, both results are consistent with the peculiar velocity signal of our fiducial $\Lambda$CDM model, ruling out the noise-only model with zero peculiar velocities at greater than $7\sigma$ (SNe) and $8\sigma$ (galaxies). When the two data sets are combined appropriately, the precision of the test increases slightly, resulting in a constraint on the signal amplitude of $A = 1.05_{-0.21}^{+0.25}$, where $A = 1$ corresponds to our fiducial model. Equivalently, we report an 11\% measurement of the product of the growth rate and amplitude of mass fluctuations evaluated at $z_\text{eff} = 0.02$, $f \sigma_8 = 0.428_{-0.045}^{+0.048}$, valid for our fiducial $\Lambda$CDM model. We explore the robustness of the results to a number of conceivable variations in the analysis and find that individual variations shift the preferred signal amplitude by less than $\sim$0.5$\sigma$. We briefly discuss our Supercal SN Ia results in comparison with our previous results using the JLA compilation.}


\begin{document}
\maketitle

\section{Introduction} \label{sec:intro}

Galaxies in the universe respond to the gravitational pull of large-scale structure, leading to the so-called peculiar velocities. This extra velocity shifts the redshift of the galaxy via the Doppler effect: $(1 + z_\text{obs}) = (1 + z)(1 + v_\parallel /c)$, where $z$ and $z_\text{obs}$ are the true and observed redshift, and $v_\parallel$ is the peculiar velocity projected along the line of sight. Peculiar velocities of galaxies are not random; roughly speaking, objects physically close to each other are being pulled by similar large-scale structures and are therefore more likely to have similar velocities. The statistical properties of the velocity field are related to the matter power spectrum and are straightforward to calculate \cite{Kaiser:1989kb,Gorski_etal}.

Measuring peculiar velocities is also, in principle, straightforward. Since an object's observed redshift is a combination of the Hubble expansion redshift and the peculiar velocity contribution, an independent estimate of the Hubble redshift is required. At low redshifts $z \ll 1$, the Hubble law applies, and we have $cz \approx H_0 d$, where $H_0$ is the Hubble constant and $d$ is the proper distance. In this limit, there is negligible dependence on cosmology (apart from $H_0$, which we will effectively marginalize over). Therefore, if one can obtain an independent distance measurement, one can estimate the peculiar velocity. This basic strategy has been employed for well over three decades \cite{Kaiser:1989kb,Gorski_etal,Sandage:1979re,Watkins:2008hf,Nusser:2011tu,Macaulay:2011av,Branchini:2012rb,Feix:2014bma}.

The challenging aspect is that peculiar velocities of $\sim$300~km/s are typically much smaller than the Hubble expansion velocity; the two are similar in size only at the very lowest redshifts, $z \sim 0.001$. The signal-to-noise ratio for the measured velocity of a single object is $v_\parallel/(cz \sigma_{\ln d})$, which is proportional to $1/z$ for a fixed fractional distance error $\sigma_{\ln d}$. For a 10\% distance measurement ($\sigma_{\ln d} = 0.1$), the velocity signal-to-noise per object is less than unity for $z > 0.01$. The study of the peculiar velocity field therefore requires the statistical power of hundreds or thousands of objects. These measurements, in turn, have the ability to constrain the cosmological model, which predicts the typical size of the velocities and their pairwise correlations. Such an approach has originally been used to constrain the matter density as well as the galaxy bias \cite{1991MNRAS.252....1K,Hudson:1995iy,Willick:1996km,Sigad:1997gc,Zaroubi:2002hh,Pike:2005tm}. More recently, the velocity measurements have been used to test for consistency with expectations from the $\Lambda$CDM model \cite{Haugboelle:2006uc,Gordon:2007zw,Ma:2010ps,Dai:2011xm,Nusser:2011tu,Weyant:2011hs,Ma:2012tt,Rathaus:2013ut,Feindt:2013pma,Ma:2013oja,Feix:2014bma,Johnson:2014kaa}, to measure cosmological parameters \cite{Abate:2008au,Turnbull:2011ty,Macaulay:2011av,Branchini:2012rb,Castro:2014oja,Carrick:2015xza}, or to test the statistical isotropy of the universe \cite{Schwarz:2007wf,Kalus:2012zu,Yang:2013gea,Appleby:2014kea,Lin:2015rza,Javanmardi:2015sfa,Bengaly:2015dza}. Others have highlighted the importance of accounting for peculiar velocities when constraining dark energy with SN Ia data \cite{Hui:2005nm,Cooray:2006ft,Neill:2007fh,Davis:2010jq} and forecasted the ability of future peculiar velocity surveys to constrain cosmology \cite{Cooray:2005yp,Hannestad:2007fb,Bhattacharya:2010cf}.

Our goal in this paper is to test the standard $\Lambda$CDM cosmological model by searching for the presence of the predicted velocities and their correlations. We will use modern peculiar velocity data and leverage the full statistical power of each individual object to perform a single test. Specifically, we define the covariance matrix of measured magnitude residuals as the sum of signal and noise contributions,
\begin{equation}
\mathbf{C} \equiv \left\langle \mathbf{\Delta m} \, \mathbf{\Delta m}^\top \right\rangle_\text{observed} = A \, \mathbf{S} + \mathbf{N} \ ,
\label{eq:Cov_intro}
\end{equation}
where $\mathbf{\Delta m}$ is the vector of magnitude residuals, which are linearly related to the peculiar velocities, while $\mathbf{S}$ and $\mathbf{N}$ are, respectively, the signal and noise covariance matrices (see section~\ref{sec:method} for definitions). Our goal then is to constrain the parameter $A$ or, equivalently, the product of the growth rate and amplitude of mass fluctuations $f\sigma_8$ which is proportional to $A^{1/2}$ (with only small dependence on other cosmological parameters). We apply formalism similar to that which we recently outlined in \cite{Huterer:2015gpa}, hereafter referred to as HSS, where we explicitly constrained the amplitude $A$. Here we analyze a new SN Ia data set featuring unprecedented photometric calibration across the various low-redshift samples. In addition, we study a large sample of galaxies from the six-degree-field galaxy survey (6dFGS) with distances derived from the fundamental plane relation. We will determine whether the amplitude $A$ preferred by the data is different from zero and consistent with unity, thus performing a powerful test of our fiducial $\Lambda$CDM model.

The rest of the paper is organized as follows. In section~\ref{sec:data}, we describe the SN and galaxy samples we use in the analysis. In section~\ref{sec:method}, we describe our methodology, which largely follows our approach in HSS. We review the calculation of the signal covariance matrix and then describe our likelihood analysis in detail. In section~\ref{sec:results}, we present the results of our test and evaluate the robustness of these results to several conceivable variations in the analysis. In section~\ref{sec:discuss}, we discuss our new results in comparison to those of previous studies and then summarize our conclusions.

\section{Data Sets} \label{sec:data}

In this section, we separately describe the selection of the SN Ia and galaxy data we use in the analysis.

\subsection{Supercal SN Ia sample} \label{sec:supercal}

SNe Ia are useful standard candles for measuring cosmological distances. After correcting their peak brightnesses for stretch (i.e.\ light-curve width, decline time) and color, each SN Ia can provide a distance measurement with roughly 7--10\% precision. While the total number of SNe observed is relatively small --- hundreds, rather than many thousands of galaxies --- the precise distance estimate makes the SNe useful for a wide variety of cosmological applications, including the study of the peculiar velocity field that is the subject of this work.

For our analysis, we consider a new ``Supercal'' compilation of SNe Ia. The Supercal sample includes data from multiple low-redshift surveys presented and analyzed in \cite{S14b}, all with photometric systems recalibrated according to \cite{Supercal} and with distance biases corrected according to \cite{ScolnicKessler16}. The sample has 50\% more SNe than the JLA sample, primarily due to the addition of the second data release of the CSP SN survey \cite{Stritzinger11} and the addition of the CfA4 survey \cite{Hicken12}. The recalibration given in \cite{Supercal} uses the relative consistency of the Pan-STARRS1 photometry over $3\pi$ steradians of the sky to tie together the photometric systems of all the low-redshift surveys. Furthermore, \cite{ScolnicKessler16} corrects for distance biases dependent on the light-curve properties of the SNe, which have a small marginalized effect on average distances but can affect distances of individual SNe by up to 0.3~mag.

Note that neither the recalibration nor the bias corrections were featured in the JLA analysis. Each of these will have some impact on inferences of $A$, as our analysis measures peculiar velocities that are coherent across the sky, and our results are thus more sensitive to biases in individual SNe or subsamples located in particular regions of the sky than the usual isotropic analysis that is suitable for measuring expansion history and dark energy parameters.

For the Supercal analysis, we employ the SALT2 light-curve fitter \cite{Guy:2007dv}, which provides a peak magnitude, stretch ($x_1$), and color ($c$) for each SN light curve, along with associated errors. Reasonable data quality cuts were applied to remove SNe which are not expected to follow the empirical standardization relations. Specifically, we keep only SNe with $x_1 < 3$, $\sigma_{x_1} < 1$, $c < 0.3$, $\sigma_c < 0.1$, a light-curve fit probability greater than $10^{-3}$, and an uncertainty in the time of peak brightness of less than two days. After a $\Lambda$CDM fit to the Hubble diagram, we apply a $4\sigma$ outlier rejection. In our main analysis, the stretch and color correction coefficients are held fixed at their best-fit values from this fit ($\alpha = 0.14$, $\beta = 3.1$). Since these parameters are well-determined from the full Hubble diagram fit and therefore measured independently of the low-redshift SNe, this simplification will not significantly affect our results.

For each SN subsample, we have included calibration systematic uncertainties by adding a correlated component to the covariance matrix following the Supercal analysis \cite{Supercal}. Therefore, the noise covariance matrix for SNe, corresponding to $\mathbf{N}$ in eq.~\eqref{eq:Cov_intro}, has non-negligible off-diagonal components. Calibration systematic uncertainties have comprised $>80\%$ of the total systematic uncertainty in past analyses (e.g.\ \cite{S14b,Betoule:2014frx}), and for the present analysis we include only these systematics. While other systematic uncertainties may have a significant impact on measurements of the dark energy equation of state due to differences between the low-$z$ and high-$z$ samples, they are likely to be much less important for an analysis of just the lowest-$z$ SNe. The calibration systematics are at the $\lesssim 2\%$ level for the different subsamples.

The final Supercal dataset contains 164 objects at $z < 0.05$ and 208 at $z < 0.1$, where the latter is the maximum redshift used in our fiducial analysis. While this sample is smaller than some low-redshift SN samples used in previous peculiar velocity studies (e.g.\ \cite{Schwarz:2007wf,Davis:2010jq,Kalus:2012zu,Ma:2012tt,Johnson:2014kaa,Appleby:2014kea}), it contains only SNe which have been placed on a consistent, and newly improved, calibration footing. Note that the Johnson et al.~\cite{Johnson:2014kaa} SN compilation consists of multiple SN samples, each with its own (and often loose) light-curve quality and parameter cuts, and each fit with either a different light-curve fitter or different reddening law. As different SN samples cover different parts of the sky, this approach could introduce large systematic biases in distance ($\sim$10\% \cite{Kessler:2009ys}) which would propagate to biases in the measurement of the velocity covariance across the sky. Such distance biases can be partially mitigated by comparing overlapping SNe in the different samples, though likely not below 5\% due to the limited statistics of overlapping SNe \cite{S14b}. Our use of a uniformly calibrated and fitted SN sample avoids these serious concerns.

\subsection{6dFGS fundamental plane sample} \label{sec:6df}

The fundamental plane (FP) describes an empirical relation \cite{Dressler:1987ny,Djorgovski:1987vx} connecting various properties of elliptical galaxies, most commonly their effective physical radius, central velocity dispersion, and average surface brightness. In the three-dimensional space of these observables, elliptical galaxies exhibit a small scatter in a particular direction and thus fall roughly on a plane, which can be written as
\begin{equation}
r = as + bi + c \ ,
\label{eq:FP}
\end{equation}
where $r$, $s$, and $i$ are, respectively, the logarithms of physical radius, velocity dispersion, and surface brightness. The parameters $a$, $b$, and $c$ are unknown \textit{a priori} and must be determined by a fit to data. While surface brightness and velocity dispersion can be directly measured, the physical radius must be inferred from the angular size. By definition, ${r = r_\text{ang} + \log_{10} d_A}$, where $d_A$ is the angular diameter distance. Fitting galaxies to the FP relation allows a determination of the radius $r$ and therefore the distance $d_A$ for each galaxy.

The six-degree-field galaxy survey (6dFGS; \cite{Jones:2004zy,Jones:2009yz}) has mapped the majority of the southern sky and obtained redshifts for over 100,000~galaxies, resulting in a 2.4$\sigma$ detection of the baryon acoustic oscillations along with a 4.5\% measurement of the distance to $z = 0.106$ \cite{Beutler:2011hx}, which is the lowest-redshift BAO distance measurement to date.

With this large sample of low-redshift galaxies, 6dFGS also allows unprecedented studies of local large-scale structure and bulk flows. A suitable subsample of 6dFGS galaxies was selected for fitting to the FP in order to estimate distances and peculiar velocities \cite{Magoulas:2012jy,Johnson:2014kaa,Springob:2014qja,Campbell:2014uia}. Distances, relative to the background expansion, were determined for a set of 8,885~galaxies in \cite{Springob:2014qja}. In their analysis, the FP was modeled as a trivariate Gaussian in the space of the FP observables. A maximum likelihood procedure was used to fit eight free parameters, three of which define the centroid of the distribution, two of which indicate the plane's orientation ($a$ and $b$ above), and three of which describe the extent of the distribution (standard deviation) in orthogonal directions \cite{Magoulas:2012jy}.

For our main analysis, we use these reported distances and their associated errors directly.\footnote{http://www.6dfgs.net/vfield/table1.txt} In section~\ref{sec:FPsys}, we perform our own fit using a simpler model for the FP in order to check the consistency of the results for different photometric bands. Note that any correlations among the distance measurements, for instance due to uncertainties in photometric calibration, are implicitly assumed to be negligible here. In our analysis, this corresponds to a diagonal noise covariance matrix $\mathbf{N}$.

\begin{figure}[t]
\centering
\includegraphics[width=0.7\textwidth]{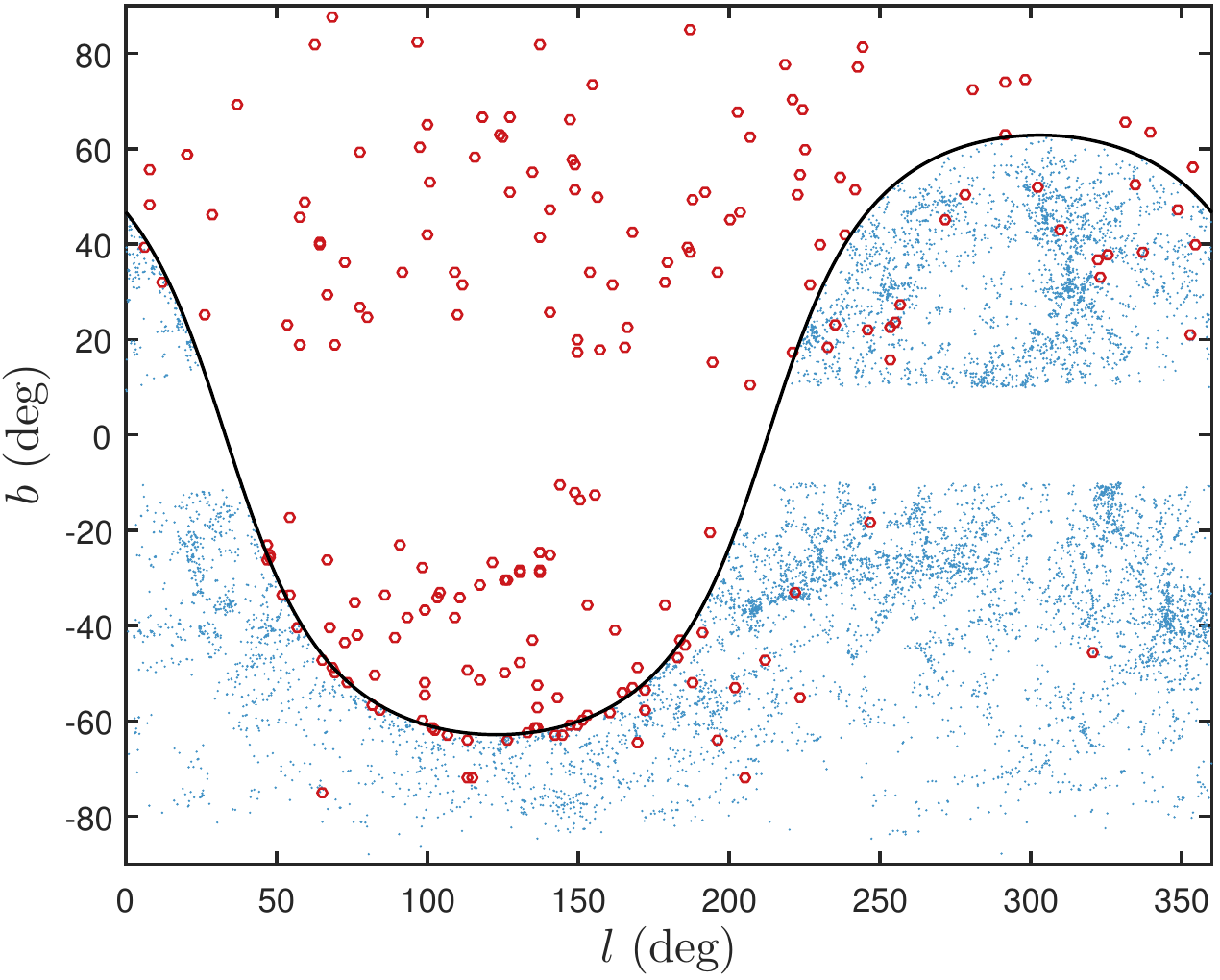}
\caption{Galactic coordinates of objects in the 6dFGS FP sample (blue points) and Supercal SN Ia compilation (red circles). For the SN Ia sample, we show only the objects with $z < 0.1$, the redshift range considered in our main analysis. The solid black curve indicates the celestial equator (zero declination).}
\label{fig:survey}
\end{figure}

\section{Methodology} \label{sec:method}

The aim of this analysis is test the standard $\Lambda$CDM model for the presence of the expected peculiar velocity signal. Our basic methodology follows that of \cite{Huterer:2015gpa}, where we introduced the dimensionless parameter $A$, which represents the amplitude of the peculiar velocity ``signal'' contribution to the full covariance matrix of distance residuals. That is, the velocity covariance is given by
\begin{equation}
\mathbf{C} = A \, \mathbf{S} + \mathbf{N} \ ,
\label{eq:Cov}
\end{equation}
where $A = 1$ for our fiducial $\Lambda$CDM model and $A = 0$ for magnitude residuals that are explained by noise alone. Here, $\mathbf{S}$ is the part of the covariance due to peculiar velocities and their correlations induced by large-scale structure, while $\mathbf{N}$ represents the statistical noise, which includes measurement uncertainty, intrinsic dispersion of the distance indicator, and any systematic uncertainties that may also induce pairwise correlations in the residuals.

In this section, we first review the calculation of the peculiar velocity signal covariance described in HSS \cite{Huterer:2015gpa}. We go on to discuss our choice of likelihood and the associated statistical analysis.

\subsection{Peculiar velocity covariance} \label{sec:signal}

The covariance of magnitude residuals is given by \cite{Kaiser:1989kb,Hui:2005nm,Huterer:2015gpa}
\begin{equation}
S_{ij} \equiv \left\langle \Delta m_i \, \Delta m_j \right\rangle  =\left[\frac{5}{\ln 10}\right]^2 \frac{(1 + z_i)^2}{H(z_i) d_L(z_i)} \, \frac{(1 + z_j)^2}{H(z_j) d_L(z_j)} \, \xi_{ij} \, ,
\label{eq:Sij}
\end{equation}
where $\xi_{ij}$ is the velocity covariance given by
\begin{align}
\xi_{ij} \equiv \xi_{ij}^\text{vel} &\equiv \langle(\mathbf{v}_i \cdot \nhat_i) (\mathbf{v}_j \cdot \nhat_j) \rangle \label{eq:xi} \\[0.2cm]
&= \frac{dD_i}{d\tau} \, \frac{dD_j}{d\tau} \int\frac{dk}{2\pi^2} P(k, a = 1) \sum_{\ell=0}^{\ell_{\rm max}} (2\ell + 1) j'_\ell(k\chi_i) j'_\ell(k\chi_j) \left [\mathcal{P}_\ell(\nhat_i \cdot \nhat_j) - \delta_{\ell 0} \right] \, , \nonumber
\end{align}
where primes denote derivatives of the Bessel functions with respect to their arguments. Here $\tau$ is the conformal time, $d\tau = dt/a$, $D_i$ is the linear growth function evaluated at redshift $z_i$, and $\chi_i = \chi(z_i)$ is the coordinate distance evaluated at that redshift. Further, $j_\ell(x)$ denotes the spherical Bessel functions, and $\mathcal{P}_\ell$ are the Legendre polynomials. The power spectrum $P(k, a)$ is evaluated using \texttt{CAMB} \cite{Lewis:1999bs} at the present epoch ($a = 1$) and assuming nonlinear theory modeled by \texttt{halofit} \cite{Smith:2002dz,Takahashi:2012em}. Including nonlinearities in $P(k)$ has an appreciable effect only for objects with small separations, including those at $z \lesssim 0.01$. Given that those nearby objects contribute appreciably to the overall constraint, the use of the fully nonlinear power spectrum is important. Note that only the first few terms in the sum over the multipoles contribute, except for objects which are nearly along the line of sight. We therefore assume $\ell_\text{max} = 20$ if $\cos(\theta) < 0.95$ and $\ell_\text{max} = 200$ otherwise, and we have verified that these choices lead to sufficiently accurate results. Eq.~\eqref{eq:xi} matches the expression from HSS \cite{Huterer:2015gpa} for the full-sky window, hence the appearance of the Kronecker delta function. We have shown in HSS that assuming a full-sky window is an excellent approximation.

To perform these calculations, and also throughout our analysis, we assume a (flat) $\Lambda$CDM model with parameters fixed to values consistent with \textit{Planck} \cite{Ade:2015xua} and other probes. That is, we fix the matter density relative to critical $\Omega_m = 0.3$, physical baryon and matter densities $\Omega_b h^2 = 0.0223$ and $\Omega_m h^2 = 0.14$, scalar spectral index $n_s = 0.965$, and amplitude of the primordial power spectrum $A_s = 2.0 \times 10^{-9}$ at $k = 0.05 \, h \text{Mpc}^{-1}$. For these choices, the derived value of the Hubble constant is $h = 0.683$ and the amplitude of mass fluctuations is $\sigma_8 = 0.80$. Note that we will effectively marginalize over $H_0$. Thus, apart from the combination $f\sigma_8$, our results only weakly depend on cosmology through the shape of the matter power spectrum controlled by $\Omega_m H_0$.

The direct evaluation of the covariance matrix for all objects in the survey is one computationally expensive part of our analysis. With $\mathcal{O}(10^4)$ objects for Supercal and 6dFGS combined, we must evaluate the expression in eqs.~\eqref{eq:Sij}--\eqref{eq:xi} a total of $\mathcal{O}(10^8)$ times. After tabulating the growth factor, power spectrum, and the spherical Bessel functions, the calculation takes about 12 hours on a modern 24-core desktop computer.

\subsection{Likelihood Analysis} \label{sec:like}

In HSS \cite{Huterer:2015gpa}, we modeled the SN Ia magnitude residuals as a multivariate Gaussian,
\begin{equation}
\mathcal{L}(A) \propto \frac{1}{\sqrt{|\mathbf{C}|}} \exp\left[-\frac{1}{2} \mathbf{\Delta m}^\intercal \mathbf{C}^{-1} \mathbf{\Delta m} \right] ,
\label{eq:like_mag}
\end{equation}
where $\mathbf{C} = A \, \mathbf{S} + \mathbf{N}$ and the elements of $\mathbf{\Delta m}$ are the magnitude residuals, $\mathbf{\Delta m}_i =
m^\text{corr}_i - m^\text{th}(z_i, \mathcal{M}, \Omega_m)$. Here, $\mathcal{M}$ is the zero-point offset in the magnitude-redshift relation, which we will return to below. However, it is actually the magnification $\mu$ that is Gaussian-distributed, since it is proportional to the large-scale peculiar velocity field at the low redshifts where the effect of lensing is unimportant:
\begin{equation}
\mu \stackrel{z \ll 1}{=} \frac{2}{a H \tilde{\chi}} \left(v_\parallel - v_{\parallel o} \right) \, .
\label{eq:mu}
\end{equation}
Because in HSS we limited our SN Ia samples to $z > 0.01$, where the peculiar velocity contribution to the redshift ($\sim$300~km/s) is $\sim$10\% or less, the first-order relation between magnitude and $\mu$,
\begin{equation}
\Delta m \approx -\frac{5}{2 \ln 10} \, \mu \, ,
\end{equation}
is sufficient, and thus $\Delta m$ is approximately Gaussian as well.

In the present analysis, we would like to use all of the newly calibrated SNe, including those at $z < 0.01$, where the signal-to-noise is the largest. At these lowest redshifts, the signal, expected to be Gaussian in $\mu$, is therefore not Gaussian in magnitude. On the other hand, the SN noise uncertainties are small enough ($\sim$7--10\%) that the noise distribution would be approximately Gaussian for either quantity. We therefore choose to model the SN velocities as Gaussian in $\mu$. In terms of the observed magnitude residuals and their covariance, the magnification and its covariance are given by
\begin{align}
\mu &= -2 \left[10^{\Delta m/5} - 1 \right] \, , \\
\mathbf{C}^{\mu \mu} &= \left(\frac{2 \ln 10}{5} \right)^2 \mathbf{C}^{\Delta m \Delta m} \, ,
\end{align}
where we have propagated the covariance at lowest order.

We therefore make slightly different choices for the SN and galaxy likelihoods, and the reasoning is as follows:
\begin{itemize}
\item For SNe Ia, the distance uncertainties due to measurement error and intrinsic scatter are relatively small ($\sim$7--10\%), so we can therefore expect the noise distribution to be approximately Gaussian in \emph{either} magnitude or magnification $\mu$. Since the peculiar velocity signal is expected to be Gaussian in $\mu$ but not in magnitude for SNe at $z < 0.01$, we use a SN Ia likelihood that is Gaussian in $\mu$.
\item For galaxies, distance uncertainties are large ($\sim$27\%), and the noise distributions have been shown \cite{Springob:2014qja} to be nearly Gaussian for log-distance residuals. Since the vast majority ($> 99\%$) of galaxies lie at $z > 0.01$, where the signal should be approximately Gaussian in \emph{either} $\mu$ or log-distance, we use a galaxy likelihood that is Gaussian in log-distance.
\end{itemize}

In deriving the SN Ia or galaxy distance estimates, one fits for empirical quantities that are not known \textit{a priori}. These include an intrinsic scatter term --- extra scatter in the astrophysical relation that is not explained by measurement error alone, as well as a constant distance offset --- the $\mathcal{M}$ parameter corresponding to the SN Ia absolute magnitude or the $c$ parameter in eq.~\eqref{eq:FP} for the galaxy FP. These ``nuisance'' parameters have already been fixed to their best-fit values in our data; however, since we are now improving the model by introducing the signal covariance, it would not be surprising if the data prefer to shift the values of these parameters. For instance, the inferred intrinsic scatter should be smaller, since we are now explaining some of this scatter with the peculiar velocity signal.

In order to avoid any potential bias, we fully marginalize over these parameters in our analysis. That is, in addition the signal amplitude $A$, we introduce $\Delta \sigma_\text{int}^2$ and $\Delta \mathcal{M}$ as parameters and let
\begin{align}
\mathbf{N} &\rightarrow \mathbf{N} + \Delta \sigma_\text{int}^2 \, \mathbf{I} \, , \\
\Delta m_i &\rightarrow \Delta m_i + \Delta \mathcal{M} \, ,
\end{align}
where $\mathbf{I}$ is the identity matrix. Here we assume flat priors on $\Delta \sigma_\text{int}^2$ and $\Delta\mathcal{M}$ and compute the likelihood over a grid of parameter values.

In addition to the separate SN Ia and galaxy analyses, we also perform a combined analysis with the hope of improving the precision of our test and our constraints on $A$. Given that the nuisance parameters --- the intrinsic scatters and distance offsets --- are unique to the separate data sets, one might be tempted to simply multiply the marginalized posteriors for $A$. However, this relies on the statistical independence of the two data sets. While there is little overlap in survey footprint between the SNe and galaxies (see figure~\ref{fig:survey}), the objects are at very low redshifts and still physically close. For the combined analysis, then, we must compute a joint signal covariance matrix and include the non-zero covariances between the SN and galaxy blocks. In other words, $S_{ij}$ is now given by eq.~\eqref{eq:Sij} with $i$ and $j$ running over both SN Ia and galaxies, while the noise covariance is block diagonal. With this in hand, we construct a combined Gaussian likelihood, employing $\mu$ as the SN observable and log-distance as the galaxy observable. Of course, we must scale the SN-galaxy covariance block by a constant factor $(2/5)\ln 10$ to account for the difference in the observable for these two types of objects.

The combined analysis requires us to vary five parameters --- $A$, plus two nuisance parameters ($\Delta \sigma_\text{int}^2$ and $\Delta\mathcal{M}$) for each data set. Since each likelihood computation involves effectively inverting a large matrix, we opt for an MCMC approach to reduce the number of likelihood evaluations. We use the basic version of the Metropolis-Hastings algorithm, and since the dimensionality of the parameter space is relatively small with most of the parameters well-constrained, convergence is not an issue. We use a Gaussian kernel (with bandwidth 0.06) to smooth the marginalized posterior distribution for $A$.

\section{Results: Constraining the signal covariance amplitude} \label{sec:results}

\begin{figure}[t]
\centering
\includegraphics[width=0.7\textwidth]{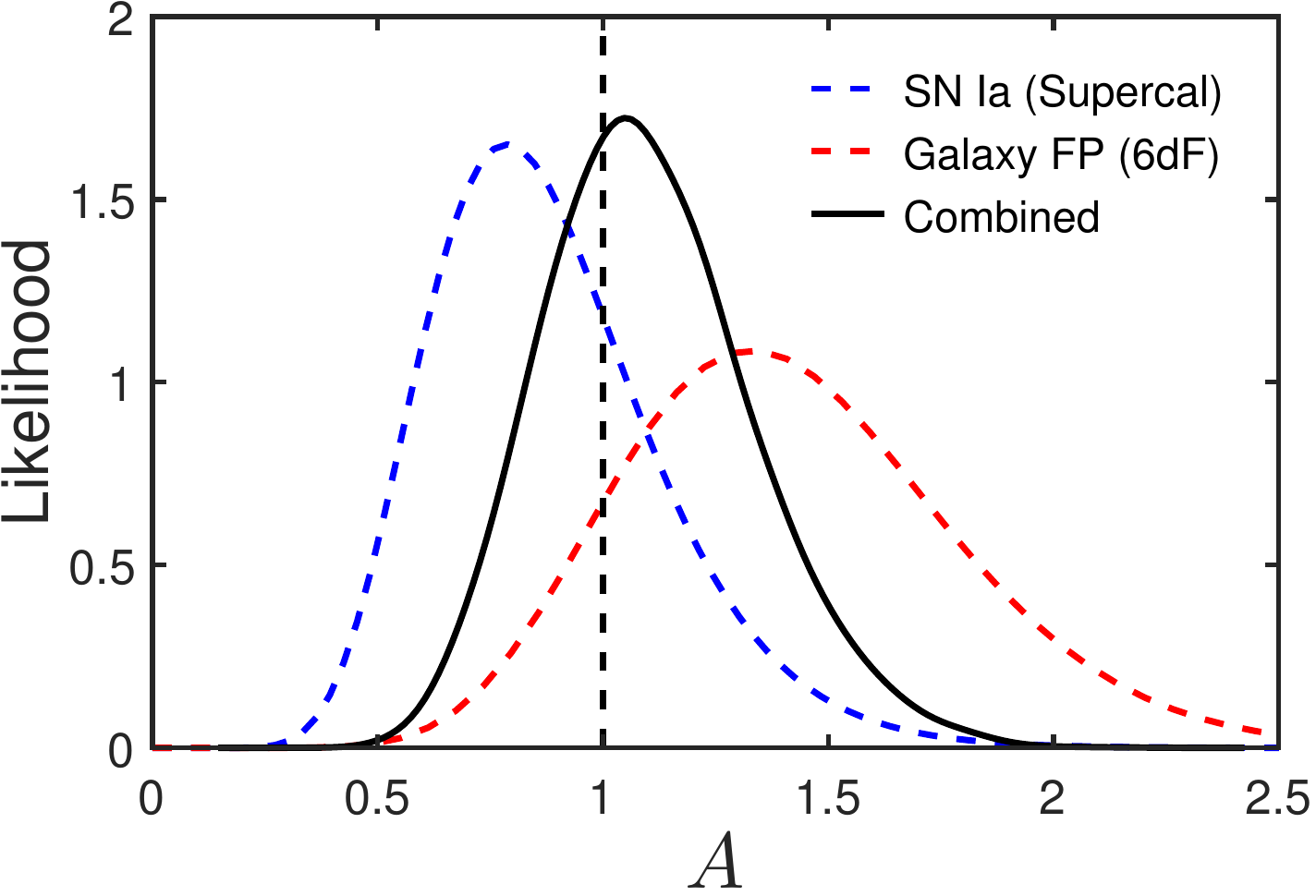}
\caption{Constraints on the amplitude of signal covariance separately for the Supercal SN Ia compilation (dashed blue), the 6dFGS FP sample (dashed red), and the combined analysis (solid black). The likelihood curves are normalized by area, and the vertical, dashed black line at $A = 1$ corresponds to our fiducial $\Lambda$CDM model.}
\label{fig:like_A}
\end{figure}

In figure~\ref{fig:like_A}, we show our results for the likelihood of $A$, the amplitude of signal covariance in the peculiar velocity data, separately for Supercal SN data (dashed blue) and 6dFGS galaxy data (dashed red). In both cases, we have marginalized over a shift in the intrinsic scatter and in the constant offset as described in section~\ref{sec:like}. We also show the results of the combined analysis (solid black), where we have marginalized over the four nuisance parameters (intrinsic scatters and offsets for both SN Ia and galaxy data).

In table~\ref{tab:summary}, we summarize numerically the constraints shown in figure~\ref{fig:like_A}. For each data set, we list the maximum-likelihood (ML) value for $A$, the 68.3\% and 95.4\% confidence intervals (CI), and the mean and standard deviation of the distribution. Note that, while Supercal and 6dFGS prefer somewhat different values for $A$, we have verified that the constraints are mutually consistent; if each were an independent measurement of $A$, the probability that the combined $\chi^2$ relative to the best-fit $A$ would be larger than what we observe, due to chance alone, is $p = 0.2$, indicating discrepancy at only $\sim$1.3$\sigma$.
  
We also list the $\Delta \chi^2$ value corresponding to $A = 0$. Although we write $\Delta \chi^2$ by convention, we are of course comparing the more general $-2 \, \Delta \ln \mathcal{L}$, which includes the term for the Gaussian prefactor in the likelihood $\mathcal{L}$. For this calculation, we minimize $\chi^2$ separately for both hypotheses, $A = 0$ and $A$ free, varying all of the nuisance parameters to avoid unfairly penalizing the $A = 0$ hypothesis. Note that this procedure is one step away from a true model comparison test for which one would include a term to penalize the model with $A$ free for having one extra free parameter. For instance, given the $\Delta \chi^2$ values in table~\ref{tab:summary}, the Akaike Information Criterion (AIC) test result is simply $\Delta \text{AIC} = \Delta \chi^2 - 2$, a small difference in our case.

\renewcommand{\arraystretch}{1.5}
\setlength{\tabcolsep}{3.5pt}
\begin{table}[b]
\centering
\begin{tabular}{|ccccccl|}
\hline
Data Set & ML & 68.3\% CI & 95.4\% CI & $\langle A \rangle \pm \sigma_A$ & $\left.\Delta\chi^2\right|_{A = 0}$ & $f \sigma_8 \, (z = 0.02)$ \\
\hline
SN Ia (Supercal) & 0.78 & (0.58, 1.06) & (0.42, 1.41) & $0.88 \pm 0.26$ & 58.4 & $0.370_{-0.053}^{+0.060}$ \\
Galaxy FP (6dFGS) & 1.33 & (1.00, 1.72) & (0.72, 2.18) & $1.42 \pm 0.37$ & 68.7 & $0.481_{-0.064}^{+0.067}$ \\
SN Ia + Galaxy & 1.05 & (0.84, 1.30) & (0.65, 1.58) & $1.10 \pm 0.24$ & 137.6 & $0.428_{-0.045}^{+0.048}$ \\
\hline
\end{tabular}
\caption{Summary of constraints on the amplitude $A$ of the signal covariance. For each data combination, we list the maximum-likelihood (ML) value, the 68.3\% and 95.4\% confidence intervals (CI), and the mean and standard deviation of the posterior distribution for $A$. We also list the $\Delta \chi^2$ value corresponding to the $A = 0$ hypothesis along with the 68.3\% CL constraint on $f \sigma_8$ ($z_\text{eff} = 0.02$) inferred from the constraint on $A$ and the fiducial model.}
\label{tab:summary}
\end{table}

\begin{figure}[t]
\centering
\includegraphics[width=0.8\textwidth]{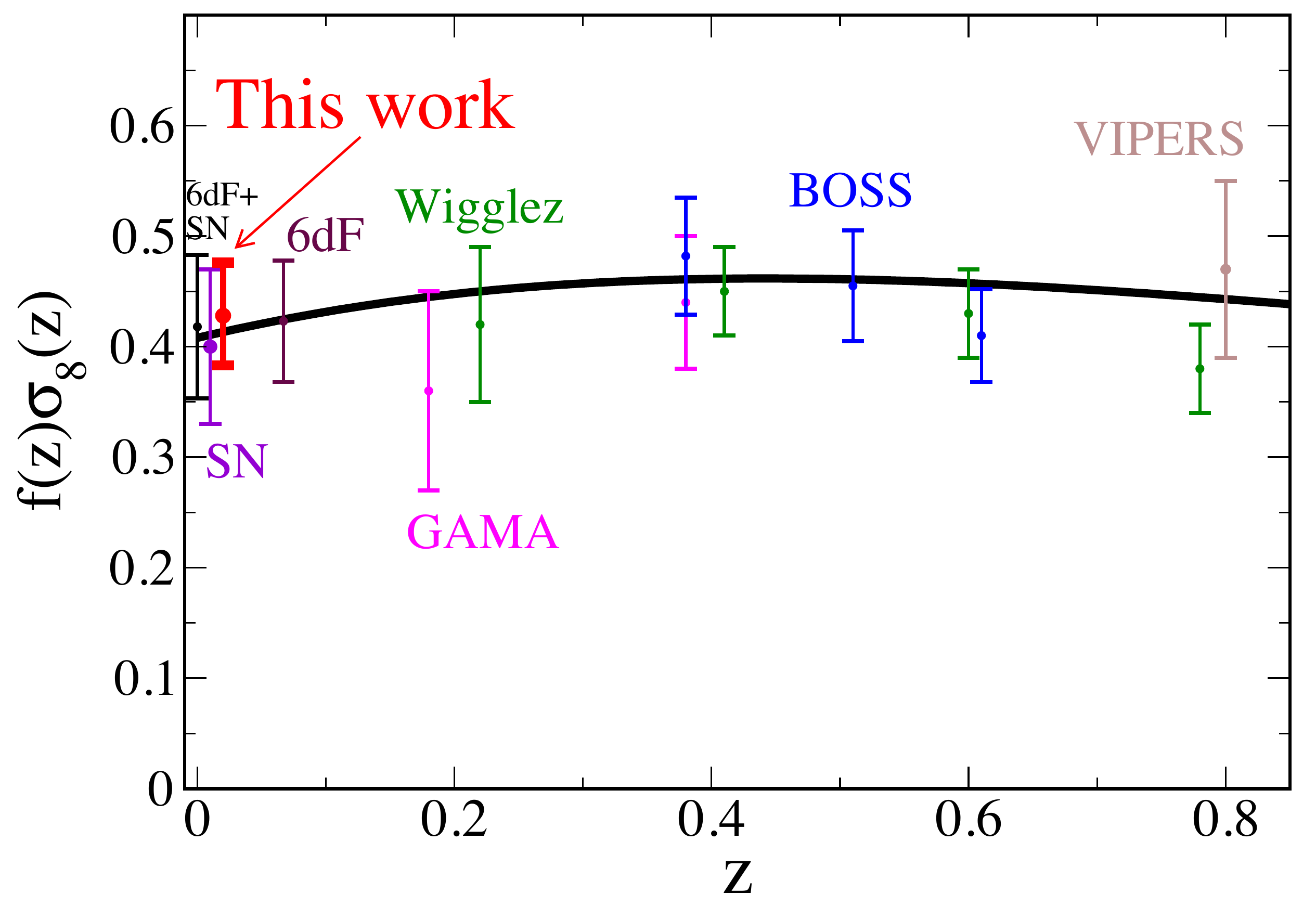}
\caption{Our constraint on the combination $f\sigma_8$, which we measure at redshift $z_\text{eff} = 0.02$ from the combined analysis of SN Ia and galaxy velocities, is shown as the red data point. We also show past measurements of the same quantity from 6dFGS at $z \simeq 0$ and SNe (black leftmost data point \cite{Johnson:2014kaa}), SNe alone (purple data point \cite{Turnbull:2011ty}), 6dFGS alone at $z = 0.067$ \cite{Beutler:2012px}, GAMA \cite{Blake:2013nif}, WiggleZ \cite{Blake:2011rj}, BOSS \cite{Beutler:2016arn}, and VIPERS \cite{delaTorre:2013rpa}. The solid line shows the prediction corresponding to our fiducial flat $\Lambda$CDM cosmology.}
\label{fig:fsig8}
\end{figure}

At leading order, the parameter $A$ is proportional to the cosmological parameter combination $(f \sigma_8)^2$. The dependence on other cosmological parameters (e.g.\ $\Omega_m$, $\Omega_b$, $n_s$) is negligible at the level of our current constraints, given variations in those parameters allowed by the \textit{Planck} data. Hence, given the value $f \sigma_8 = 0.418$ derived for our fiducial $\Lambda$CDM model at the effective redshift $z_\text{eff} = 0.02$, we can easily convert the constraints on $A$ into constraints on $f \sigma_8$ ($z_\text{eff} = 0.02$), which are also listed in table~\ref{tab:summary}. Note that the effective redshift of our constraint, taken to be the $(S/N)^2$-weighted mean redshift, is 0.014 for the SN sample and 0.024 for the galaxy sample, though for simplicity, we quote the constraint at $z_\text{eff} = 0.02$ in all cases, as the difference is negligible. Note that here we are following most literature on the subject and only varying the combination $f \sigma_8$, fixing all other cosmological parameters (e.g.\ $\Omega_m$, $\Omega_b$, $n_s$) to fiducial values. This is a reasonably good approximation, given that the velocity covariance signal largely depends on $f \sigma_8$, but we note that a full analysis, one which combines constraints from multiple cosmological probes, would involve simultaneously varying all cosmological parameters.

As apparent from figure~\ref{fig:like_A} and table~\ref{tab:summary}, we find the data consistent with $A = 1$ at $1\sigma$ (68.3\% CL) in all cases. The $A = 0$ hypothesis is ruled out at $> 7\sigma$ by SN Ia data and $> 8\sigma$ by galaxy data.

As discussed in section~\ref{sec:like}, the combined analysis is not trivial in the sense that one cannot simply multiply the individual SN Ia and galaxy posteriors because there are significant peculiar velocity signal correlations between SN and galaxy pairs. Instead, we proceed as described in section~\ref{sec:like} and include the cosmological correlations between individual SN and galaxy velocities, which depend on the angular positions and redshifts of the objects. We find that combining the two sets moderately improves the precision of our test, and we obtain the constraint $A = 1.05^{+0.25}_{-0.21}$ at 68.3\% confidence. The $A = 0$ hypothesis is ruled out at $> 11\sigma$, and the standard error of $A$ is reduced by roughly 30\% relative to that for galaxies, though only slightly relative to that for SNe.

Scaling from our fiducial model, we convert our constraint on $A$ into a constraint on $f \sigma_8 \, (z_\text{eff} = 0.02) = 0.428^{+0.048}_{-0.045}$ for the combined analysis of SNe and galaxies. In figure~\ref{fig:fsig8}, we show this constraint along with other constraints on this parameter combination from major galaxy surveys over the past decade. We see that our constraint is very competitive with, and complementary to, the other existing constraints.

\subsection{Robustness of the SN Ia analysis} \label{sec:SNsys}

One might wonder whether our choice to treat the SN Ia residuals as Gaussian in magnification $\mu$, which is proportional to the velocities, rather than Gaussian in magnitude (logarithm of distance), has an appreciable effect on our results.

\begin{figure}[t]
\centering
\includegraphics[width=0.470\textwidth]{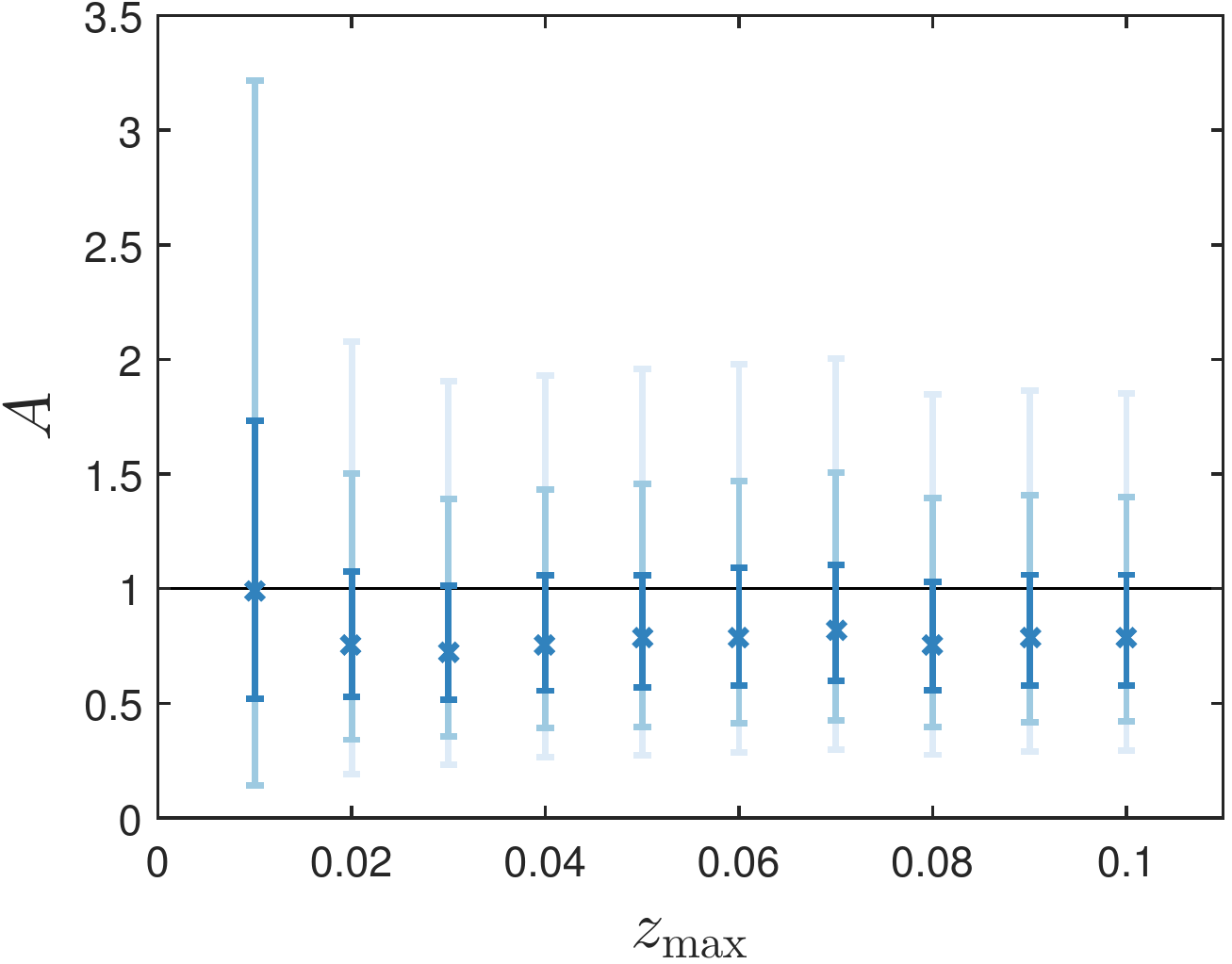}
\hspace{0.03\textwidth}
\includegraphics[width=0.485\textwidth]{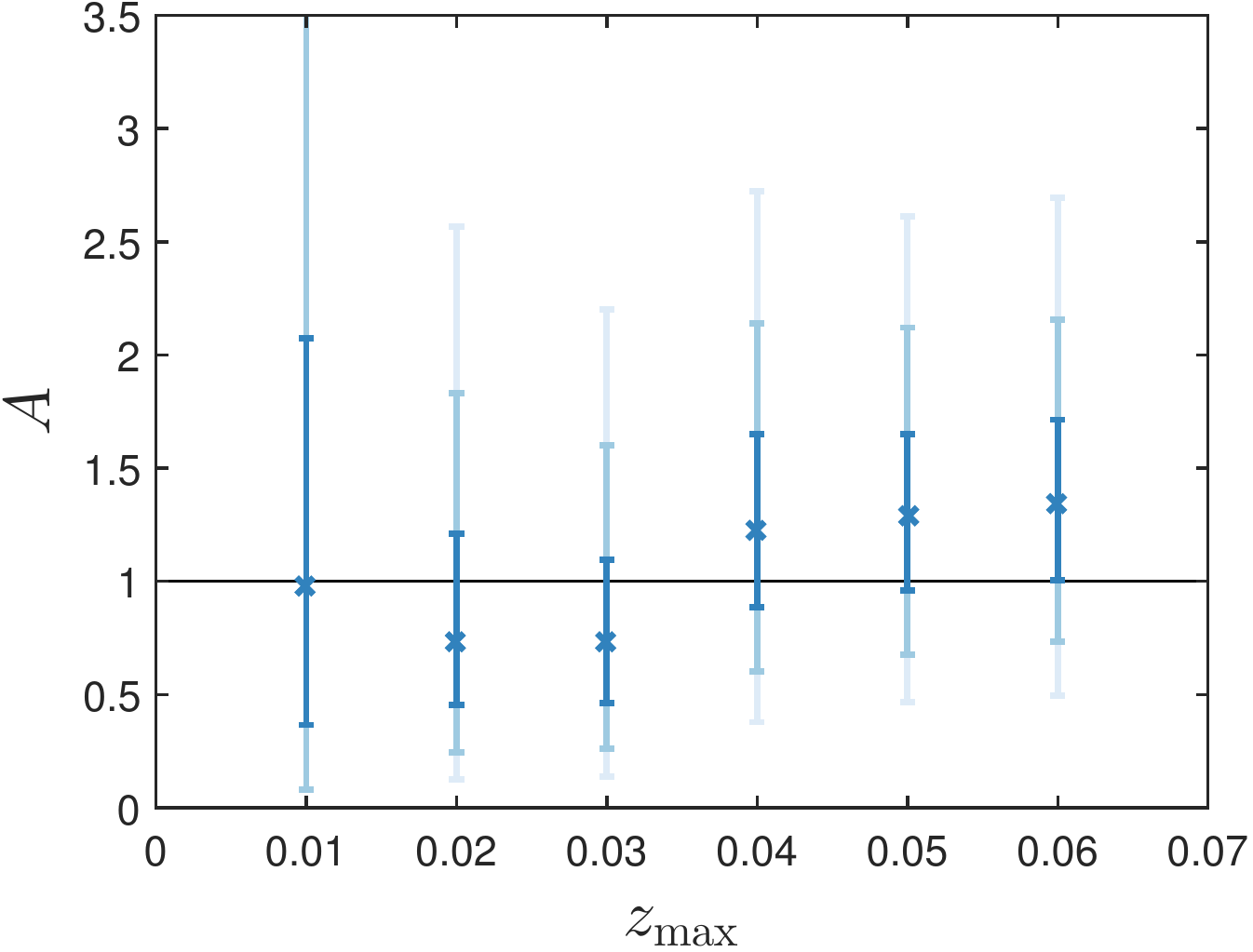}
\caption{Constraints on the amplitude of signal covariance as a function of the maximum redshift used in the analysis for SNe Ia (left panel) and galaxies (right panel). The overlapping error bars denote the 68.3\%, 95.4\%, and 99.7\% confidence limits for a given $z_\text{max}$.}
\label{fig:zmax}
\end{figure}

First, as a sanity check, we restrict the sample to SNe with $z > 0.01$, where the peculiar velocity contribution to the redshift ($\sim$300~km/s) is less than $\sim$10\%. Since the noise uncertainties on SN distances are also small (roughly 7--10\%), we would expect the likelihood to be approximately Gaussian in both $\mu$ and magnitude, and so our constraint on $A$ should be unaffected by this choice. We perform this check and find that, while the constraints are now weakened without the high signal-to-noise SNe at $z < 0.01$, the posterior for $A$ is nearly identical for either choice of the likelihood function.

Next, we perform the same test but include all of the SNe (up to $z = 0.1$). Using a likelihood that is Gaussian in magnitude rather than $\mu$ shifts the peak of the marginalized posterior for $A$ to 0.71, a shift of $-0.07$ or $\sim$0.3$\sigma$. The mean value is similarly shifted, while the uncertainty is not significantly changed. This variation therefore produces changes in $A$ comfortably smaller than the statistical error. Furthermore, given the linear relation between $\mu$ and velocity (eq.~\eqref{eq:mu}), we expect a likelihood that is Gaussian in $\mu$ to be much closer to the true likelihood and, correspondingly, any systematic effect resulting from our choice of approximate likelihood to be smaller than this shift.

Finally, as a further check for possible systematic effects in the data, we repeat the SN analysis but vary the maximum redshift. For each $z_\text{max}$ in the left panel of figure~\ref{fig:zmax}, we show the constraints on $A$ after marginalizing over the two nuisance parameters. The results are very consistent as we vary $z_\text{max}$. It is also interesting to note that the constraints negligibly improve after $z_\text{max} \simeq 0.05$ and remain unchanged after $z = 0.1$, illustrating the importance the lowest-redshift SNe. On the other hand, the handful of SNe at $z < 0.01$ cannot provide useful constraints by themselves, particularly if we expect them to constrain the two nuisance parameters as well.

We thus find no evidence for any systematic effects that contribute significantly in comparison to the statistical uncertainty in $A$. The fact that SNe furnish such a precise distance estimate and have well-studied systematics suggests they will continue to be a useful probe of peculiar velocities, even if they are relatively much smaller in number than galaxies.

\begin{figure}[t]
\centering
\includegraphics[width=0.485\textwidth]{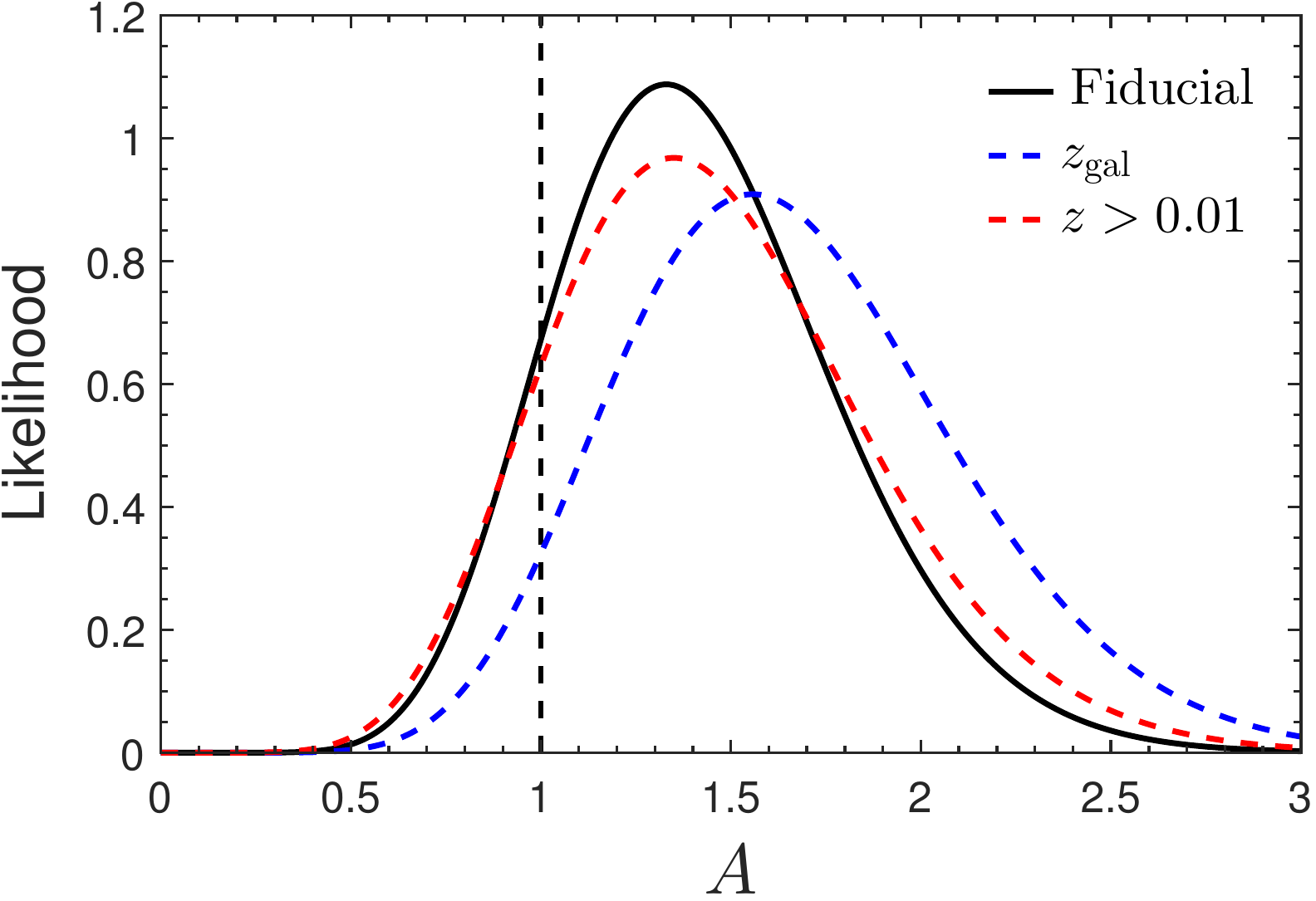}
\hspace{0.015\textwidth}
\includegraphics[width=0.485\textwidth]{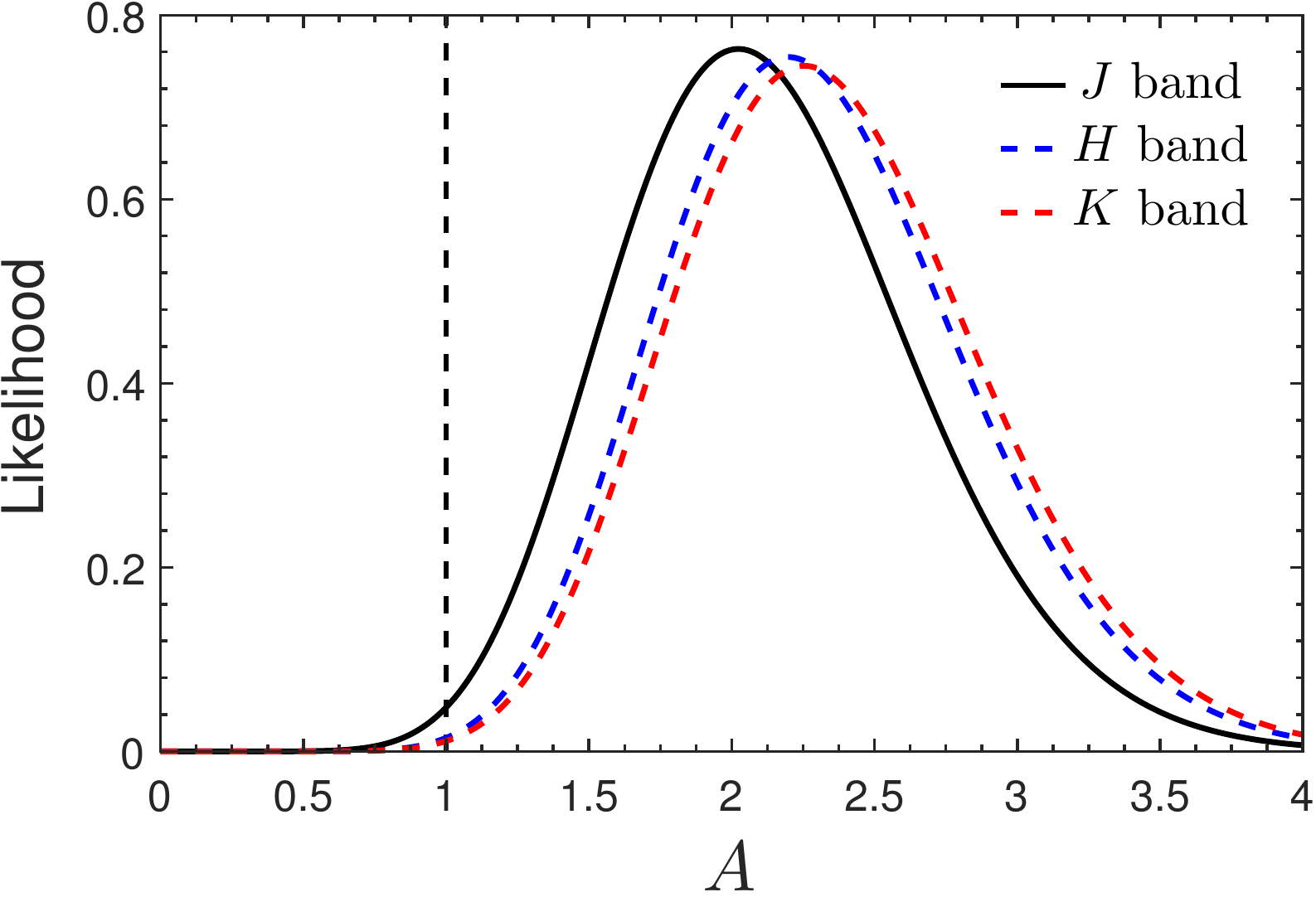}
\caption{Effect of variations in the galaxy velocity analysis on $A$ constraints. In the left panel, the fiducial analysis (solid black curve) is compared to variations with the redshift range restricted to $z > 0.01$ (dashed red) and with galaxy redshifts used in place of any estimated group/cluster redshifts (dashed blue). The right panel shows constraints on $A$ for alternative fits to the FP using orthogonal regression under the assumption of an infinite plane with uniform intrinsic scatter. We compare results for the $J$-band (solid black), $H$-band (dashed blue), and $K$-band (dashed red) photometry.}
\label{fig:Asys}
\end{figure}

\subsection{Robustness of the galaxy FP analysis} \label{sec:FPsys}

As illustrated in figure~\ref{fig:like_A} and shown in table~\ref{tab:summary}, our nominal constraints on $A$ from 6dFGS galaxy data are consistent with $A = 1$ and rule out the $A = 0$ model at $> 8\sigma$. Here we will explore the robustness of this result to a number of conceivable variations in the fiducial analysis.

As discussed in section~\ref{sec:like}, at the lowest redshifts ($z \lesssim 0.01$) where the peculiar velocities are comparable to the cosmological redshift, the signal is expected to be approximately Gaussian in the velocities and magnification and therefore cannot be Gaussian in terms of magnitude or log-distance. Since the FP-derived galaxy distances have (noise) distributions that are nearly Gaussian in log-distance, and since the vast majority of galaxies lie at $z > 0.01$ where the signal should be approximately Gaussian in either quantity, we have chosen to model the galaxy velocities as Gaussian in log-distance. To investigate the extent to which this choice may bias our result, we simply repeat the analysis without the $z < 0.01$ galaxies altogether. The corresponding constraints on $A$ are shown in the left panel of figure~\ref{fig:Asys}; they are nearly identical to the constraints from the fiducial analysis and only slightly weaker. This illustrates that, not only are the lowest-redshift galaxies not biasing the result, they do not contribute significantly to the constraint on $A$.

In our fiducial analysis, we default to using the group/cluster redshifts estimated for galaxies that have been identified as group/cluster members, and indeed the 6dFGS galaxy distances were derived assuming these redshifts. To test whether this choice affects our results, we repeat the analysis using the individual galaxy redshifts throughout. The constraints on $A$ for this scenario are also shown in the left panel of figure~\ref{fig:Asys} (the curve labeled $z_\text{gal}$). We find that using the individual galaxy redshifts moderately weakens the constraints and shifts the peak of the likelihood by 0.22 to $A = 1.55$, a shift of $\sim$0.5$\sigma$ (relative to these weaker $z_\text{gal}$ constraints).

Finally, we would like to investigate the extent to which observational systematics, or systematics related to the FP relation, may affect our results. The galaxy distances we adopt were estimated by fitting the $J$-band FP using a trivariate Gaussian model. Although distances were derived for $J$-band photometry only, there are also velocity dispersion and surface brightness measurements derived from observations in the $H$ and $K$ bands. Since the FP is an empirical relation, and since there is no fundamental reason why the $J$ band should be used, one might wonder whether cosmological results from the other bands are consistent. After all, the FP is fit empirically, with substantial astrophysics in play, and it is not hard to imagine that different bands may be affected by different astrophysical systematics.

To study the extent to which such systematics may affect our results, we re-fit the FP ourselves for all three bands using the FP data described in \cite{Campbell:2014uia}. The procedure used in \cite{Springob:2014qja} for fitting the trivariate Gaussian model is rather involved, so we adopt a simpler but common approach: treat the FP as an infinite plane with uniform scatter and fit the plane using a type of orthogonal regression. Similar fits were performed in \cite{Hyde:2008yh,LaBarbera:2009vn,Robotham:2015fia}, among others. We adopt the likelihood for multidimensional orthogonal regression that is described well in \cite{Robotham:2015fia}, and we use some of their notation. Up to an irrelevant additive constant, we have
\begin{equation}
-2 \ln \mathcal{L}_\text{FP} = \sum_i w_i \left[\frac{\left(\bm{\alpha}^\intercal \bm{x}_i + c \right)^2}{\sigma_i^2} + \ln(\sigma_i^2) - \ln(\bm{\alpha}^\intercal \bm{\alpha}) \right] \, ,
\end{equation}
where $\bm{\alpha}^\top = [a, \ b, \ -1]$ and $\bm{x}^\top_i = [s_i, \ i_i, \ r_i]$. Note that $r$ was inferred from $r_\text{ang}$ using the same fiducial expansion history (flat $\Lambda$CDM with $\Omega_m = 0.3$) that we assume in our main analysis. The weights $w_i$ are inversely proportional to the selection probability and were computed according to \cite{Magoulas:2012jy}. The uncertainty $\sigma_i$ is given by
\begin{equation}
\sigma_i^2 = \bm{\alpha}^\top \mathbf{\Sigma}_i \, \bm{\alpha} + \sigma_r^2 \, ,
\end{equation}
where $\sigma_r$ is the intrinsic scatter about the relation projected in the $r$ direction and $\mathbf{\Sigma}_i$ is the covariance matrix for the $i^\text{th}$ galaxy's observables $s$, $i$, and $r$. Note that, as explained in \cite{Magoulas:2012jy}, the errors given for $i$ and $r$ are strongly correlated with correlation coefficient $-0.95$, and accounting for this correlation reduces the scatter $\sigma_r$ needed to explain the data.

Using the MCMC approach, we constrain the four free parameters ($a$, $b$, $c$, $\sigma_r$) of the FP. The results are shown in table~\ref{tab:FPfits}, where we list the mean and standard deviation for each parameter and for each photometric band (the posteriors are nearly Gaussian).

In the right panel of figure~\ref{fig:Asys}, we compare the constraints on $A$ assuming the infinite-plane, uniform-scatter model for the FP and using our constraints on the parameters. Overall, these constraints favor a higher amplitude of signal covariance than our main results from the FP model of \cite{Springob:2014qja}. Since this alternative FP model is embedded as a special case of their more general Gaussian model, we emphasize that this large shift is \emph{not} itself evidence of a systematic effect in our main result, though it does underscore the need to rigorously fit the empirical FP relation.

On the other hand, we can now study the effect of fitting the FP using the different photometric bands. We find that results from the $J$, $H$, and $K$ bands are in remarkable agreement, and we can estimate the size of a systematic error due to the photometry by computing the (sample) standard deviation of the three ML values for $A$ (2.02, 2.20, and 2.25). This suggests that the uncertainty is less than 0.12, comfortably smaller than our statistical uncertainty for either model of the FP.

\renewcommand{\arraystretch}{1.5}
\setlength{\tabcolsep}{4pt}
\begin{table}[t]
\centering
\begin{tabular}{|c|cc|cc|cc|}
\hline
& \multicolumn{2}{c|}{$J$ Band} & \multicolumn{2}{c|}{$H$ Band} & \multicolumn{2}{c|}{$K$ Band} \\
& ML & $\mu \pm \sigma$ & ML & $\mu \pm \sigma$ & ML & $\mu \pm \sigma$ \\
\hline
a & 1.513 & 1.513 $\pm$ 0.013 & 1.494 & 1.494 $\pm$ 0.013 & 1.492 & 1.492 $\pm$ 0.013\\
b & $-$0.8566 & $-$0.8566 $\pm$ 0.0046 & $-$0.8448 & $-$0.8448 $\pm$ 0.0044 & $-$0.8199 & $-$0.8199 $\pm$ 0.0041\\
c & $-$0.421 & $-$0.422 $\pm$ 0.030 & $-$0.293 & $-$0.294 $\pm$ 0.031 & $-$0.323 & $-$0.324 $\pm$ 0.030\\
$\sigma_r$ & 0.0885 & 0.0885 $\pm$ 0.0010 & 0.0887 & 0.0887 $\pm$ 0.0010 & 0.0865 & 0.0866 $\pm$ 0.0010\\
\hline
\end{tabular}
\caption{Fits to the FP for the 6dFGS sample under the assumption of an infinite plane with uniform scatter. The FP is fit separately for each of the three photometric bands, and in each case we list the ML values, means, and standard deviations for the inferred FP parameters.}
\label{tab:FPfits}
\end{table}

\section{Discussion and conclusions} \label{sec:discuss}

\subsection{Comparison to JLA results} \label{sec:compare}

Our results are in excellent agreement with expectations from the standard $\Lambda$CDM model; however, it is instructive to compare our SN results more carefully with those from HSS. At face value, the results here are substantially different from those obtained for the JLA sample in HSS, where we found that JLA prefers a ML value of $A = 0.19$; however, the results were found to be consistent with $A = 1$ at the 95.4\% (2$\sigma$) CL.\footnote{In HSS we also found that the Union2 sample \cite{Amanullah:2010vv} prefers $A \approx 1$, clearly in agreement with the present results using the Supercal sample.} In contrast, our present results using the Supercal sample favor the value $A = 0.78$. Quantitatively, the discrepancy between the JLA and Supercal posteriors for $A$ is not especially significant, with a p-value of $\sim$0.1. Nevertheless, it seems prudent to briefly investigate why the JLA and Supercal samples give different results.

We first select objects that are common to the two samples; at $z < 0.1$, this is a sample of 87~SNe. Using only this overlap sample, the likelihood for $A$ is broader than that from either JLA or Supercal, as expected, and for Supercal it peaks at $A = 0.42$, closer to the best-fit from JLA. We then select and fit the remaining 121~SNe at $z < 0.1$ that are unique to Supercal, leading to a likelihood with a peak at $A = 0.86$. Clearly, it is these SNe found in Supercal but not JLA that dominate the constraint and lead to a strong preference for a higher value of $A$. This is not too surprising, as the highest signal-to-noise SNe at $z < 0.01$ are not included in the JLA sample.

\subsection{Conclusions} \label{sec:conclude}

In this study, we have used redshift and distance measurements from both the Supercal SN Ia analysis and the 6dFGS FP analysis to search for the presence of peculiar velocities and their correlations predicted by the standard cosmological model. We applied the basic formalism and approach described in HSS \cite{Huterer:2015gpa}, which is particularly transparent and straightforward to implement. We used the data to constrain a single parameter of interest, the overall amplitude $A$ of the signal covariance matrix, where $A = 1$ is the value expected based on our fiducial $\Lambda$CDM model. In the analysis, we paid special attention to the modeling of the data, justifying our specific choices for the likelihood function and explicitly marginalizing over nuisance parameters (scatter, distance offsets) to avoid a potential bias.

Our results (figure~\ref{fig:like_A}, table~\ref{tab:summary}) indicate good mutual agreement between the SN and galaxy samples as well as agreement with the peculiar velocity signal of the fiducial model ($A = 1$) at $< 1\sigma$. Combining the two data sets, we obtain $A = 1.05^{+0.25}_{-0.21}$ (68.3\% CL) and rule out the zero-peculiar-velocity case ($A = 0$) at $> 11\sigma$. Equivalently, we report an 11\% measurement of the product of the growth rate and amplitude of mass fluctuations $f \sigma_8 = 0.428_{-0.045}^{+0.048}$ at an effective redshift $z_\text{eff} = 0.02$. Note that this constraint assumes $\Lambda$CDM, with other cosmological parameters (e.g.\ $\Omega_m$) fixed to fiducial values.

Our analysis is most similar to that of \cite{Johnson:2014kaa}, which also studies SN Ia and 6dFGS galaxy velocities and finds qualitatively similar results. The principal difference in the data is the SN Ia sample. Our Supercal sample, while somewhat smaller, features dramatically improved photometric calibration, with the photometric systems of different low-redshift surveys tied together consistently. Our sample selection therefore largely circumvents serious concerns about the SN Ia calibration heterogeneity in previous work (see section~\ref{sec:supercal}). Our approach is also somewhat different. Instead of separating the velocity constraints into wavenumber bins, or using binning to smooth the velocity field, we treat each object individually and constrain the amplitude of the signal covariance directly to constrain the cosmological model. In this sense, our study is complementary to that of \cite{Johnson:2014kaa}, and we note our overall agreement.

The fact that our results using a somewhat different approach are in agreement with those of previous studies suggests that peculiar velocities are finally becoming a reliable probe of large-scale structure. This is a very encouraging development relative to the state of the field in the 1990s, when peculiar velocities seemed to favor high values of the matter density now known to be incorrect (e.g.\ \cite{1993ApJ...405..437N,Bernardeau:1994vz,Dekel:1993hq}). Nevertheless, careful inspection of recent results shows that some caution is still in order, as the constraints from velocities still show dependence on both the choice of data and the analysis. For example, our Supercal results are mildly discrepant with the JLA results in HSS (though the two are formally concordant at well within 2$\sigma$). And while we argue that the Supercal analysis provides the most carefully calibrated and fitted SN sample to date, we and others still benefit from ``knowing the right answer'' for cosmological parameter values, thanks to precise cosmological constraints from the CMB and other probes. To circumvent this problem, one should introduce blinding in these analyses to avoid a subjective bias, much like the procedures routinely applied in particle physics experiments \cite{2005ARNPS..55..141K}.

In conclusion, our 11\% measurement of $f\sigma_8$ at $z_\text{eff} = 0.02$ is in excellent agreement with the prediction of the currently favored $\Lambda$CDM cosmological model. Upcoming velocity surveys such as TAIPAN\footnote{http://www.taipan-survey.org}, Widefield ASKAP L-band Legacy All-sky Blind Survey (WALLABY\footnote{http://www.atnf.csiro.au/research/WALLABY}; \cite{Johnston:2007yh,Koribalski:2012pf}) and Westerbork Northern Sky HI survey (WNSHS; see \cite{Duffy:2012gr}) will significantly increase the sample of nearby galaxies and enable a $\sim$3\% measurement of $f\sigma_8$ \cite{Koda:2013eya,Howlett:2016urc}. These constraints will complement those from redshift-space distortions at higher redshifts (see figure~\ref{fig:fsig8}), significantly extending the lever arm in redshift for constraints on dark energy and gravity and ushering in an era of precise tests of structure formation at redshifts near zero.

\acknowledgments
We thank Alex~Barreira, Chris~Blake, and Simon~White for useful discussions. DH is supported by NSF under contract AST-0807564 and DOE under contract DE-FG02-95ER40899. DH also thanks the Aspen Center for Physics, which is supported by NSF Grant No.\ 1066293, for hospitality while some of this work was carried out. D. Scolnic gratefully acknowledges support from NASA grant 14-WPS14-0048, from the Hubble Fellowship awarded by the Space Telescope Science Institute, and from the Kavli Institute for Cosmological Physics at the University of Chicago, which is supported by grant NSF PHY-1125897 and an endowment from the Kavli Foundation. FS acknowledges support from the Marie Curie Career Integration Grant (FP7-PEOPLE-2013-CIG) ``FundPhysicsAndLSS'' and Starting Grant (ERC-2015-STG 678652) ``GrInflaGal'' from the European Research Council.

\bibliography{fpvel}

\providecommand{\href}[2]{#2}\begingroup\raggedright\begin{thebibliography}{10}

\bibitem{Kaiser:1989kb}
N.~Kaiser, {\it {Theoretical implications of deviations from Hubble flow}},
  {\em Mon. Not. Roy. Astron. Soc.} {\bf 231} (1989) 149.

\bibitem{Gorski_etal}
K.~M. {Gorski}, M.~{Davis}, M.~A. {Strauss}, S.~D.~M. {White}, and A.~{Yahil},
  {\it {Cosmological velocity correlations - Observations and model
  predictions}},  {\em Astrophys. J.} {\bf 344} (1989) 1--19.

\bibitem{Sandage:1979re}
A.~Sandage, G.~A. Tammann, and A.~Yahil, {\it {The velocity field of bright
  nearby galaxies. I. The variation of mean absolute magnitude with redshift f
  or galaxies in a magnitude-limited sample}},  {\em Astrophys. J.} {\bf 232}
  (1979) 352--364.

\bibitem{Watkins:2008hf}
R.~Watkins, H.~A. Feldman, and M.~J. Hudson, {\it {Consistently Large Cosmic
  Flows on Scales of 100 Mpc/h: a Challenge for the Standard LCDM Cosmology}},
  {\em Mon. Not. Roy. Astron. Soc.} {\bf 392} (2009) 743--756,
  [\href{http://arxiv.org/abs/0809.4041}{{\tt arXiv:0809.4041}}].

\bibitem{Nusser:2011tu}
A.~Nusser and M.~Davis, {\it {The cosmological bulk flow: consistency with
  $\Lambda$CDM and $z\approx 0$ constraints on $\sigma_8$ and $\gamma$}},  {\em
  Astrophys. J.} {\bf 736} (2011) 93,
  [\href{http://arxiv.org/abs/1101.1650}{{\tt arXiv:1101.1650}}].

\bibitem{Macaulay:2011av}
E.~Macaulay, H.~A. Feldman, P.~G. Ferreira, A.~H. Jaffe, S.~Agarwal, M.~J.
  Hudson, and R.~Watkins, {\it {Power Spectrum Estimation from Peculiar
  Velocity Catalogues}},  {\em Mon. Not. Roy. Astron. Soc.} {\bf 425} (2012)
  1709--1717, [\href{http://arxiv.org/abs/1111.3338}{{\tt arXiv:1111.3338}}].

\bibitem{Branchini:2012rb}
E.~Branchini, M.~Davis, and A.~Nusser, {\it {The velocity field of 2MRS
  Ks=11.75 galaxies: constraints on beta and bulk flow from the luminosity
  function}},  {\em Mon. Not. Roy. Astron. Soc.} {\bf 424} (2012) 472--481,
  [\href{http://arxiv.org/abs/1202.5206}{{\tt arXiv:1202.5206}}].

\bibitem{Feix:2014bma}
M.~Feix, A.~Nusser, and E.~Branchini, {\it {Tracing the cosmic velocity field
  at $z \thicksim 0.1$ from galaxy luminosities in the SDSS DR7}},  {\em JCAP}
  {\bf 1409} (2014) 019, [\href{http://arxiv.org/abs/1405.6710}{{\tt
  arXiv:1405.6710}}].

\bibitem{1991MNRAS.252....1K}
N.~{Kaiser}, G.~{Efstathiou}, W.~{Saunders}, R.~{Ellis}, C.~{Frenk},
  A.~{Lawrence}, and M.~{Rowan-Robinson}, {\it {The large-scale distribution of
  IRAS galaxies and the predicted peculiar velocity field}},  {\em MNRAS} {\bf
  252} (Sept., 1991) 1--12.

\bibitem{Hudson:1995iy}
M.~J. Hudson, A.~Dekel, S.~Courteau, S.~M. Faber, and J.~A. Willick, {\it
  {Omega and biasing from optical galaxies versus POTENT mass}},  {\em Mon.
  Not. Roy. Astron. Soc.} {\bf 274} (1995) 305,
  [\href{http://arxiv.org/abs/astro-ph/9501074}{{\tt astro-ph/9501074}}].

\bibitem{Willick:1996km}
J.~A. Willick, M.~A. Strauss, A.~Dekel, and T.~Kolatt, {\it {Maximum-likelihood
  comparisons of tully-fisher and redshift data: constraints on omega and
  biasing}},  {\em Astrophys. J.} {\bf 486} (1997) 629,
  [\href{http://arxiv.org/abs/astro-ph/9612240}{{\tt astro-ph/9612240}}].

\bibitem{Sigad:1997gc}
Y.~Sigad, A.~Eldar, A.~Dekel, M.~A. Strauss, and A.~Yahil, {\it {Iras versus
  potent density fields on large scales: biasing and omega}},  {\em Astrophys.
  J.} {\bf 495} (1998) 516, [\href{http://arxiv.org/abs/astro-ph/9708141}{{\tt
  astro-ph/9708141}}].

\bibitem{Zaroubi:2002hh}
S.~Zaroubi, E.~Branchini, Y.~Hoffman, and L.~Nicolaci~da Costa, {\it
  {Consistent beta values from density-density and velocity-velocity
  comparisons}},  {\em Mon. Not. Roy. Astron. Soc.} {\bf 336} (2002) 1234,
  [\href{http://arxiv.org/abs/astro-ph/0207356}{{\tt astro-ph/0207356}}].

\bibitem{Pike:2005tm}
R.~W. Pike and M.~J. Hudson, {\it {Cosmological parameters from the comparison
  of the 2mass gravity field with peculiar velocity surveys}},  {\em Astrophys.
  J.} {\bf 635} (2005) 11--21,
  [\href{http://arxiv.org/abs/astro-ph/0511012}{{\tt astro-ph/0511012}}].

\bibitem{Haugboelle:2006uc}
T.~Haugboelle, S.~Hannestad, B.~Thomsen, J.~Fynbo, J.~Sollerman, and S.~Jha,
  {\it {The Velocity Field of the Local Universe from Measurements of Type Ia
  Supernovae}},  {\em Astrophys. J.} {\bf 661} (2007) 650--659,
  [\href{http://arxiv.org/abs/astro-ph/0612137}{{\tt astro-ph/0612137}}].

\bibitem{Gordon:2007zw}
C.~Gordon, K.~Land, and A.~Slosar, {\it {Cosmological Constraints from Type Ia
  Supernovae Peculiar Velocity Measurements}},  {\em Phys. Rev. Lett.} {\bf 99}
  (2007) 081301, [\href{http://arxiv.org/abs/0705.1718}{{\tt
  arXiv:0705.1718}}].

\bibitem{Ma:2010ps}
Y.-Z. Ma, C.~Gordon, and H.~A. Feldman, {\it {The peculiar velocity field:
  constraining the tilt of the Universe}},  {\em Phys. Rev.} {\bf D83} (2011)
  103002, [\href{http://arxiv.org/abs/1010.4276}{{\tt arXiv:1010.4276}}].

\bibitem{Dai:2011xm}
D.-C. Dai, W.~H. Kinney, and D.~Stojkovic, {\it {Measuring the cosmological
  bulk flow using the peculiar velocities of supernovae}},  {\em JCAP} {\bf
  1104} (2011) 015, [\href{http://arxiv.org/abs/1102.0800}{{\tt
  arXiv:1102.0800}}].

\bibitem{Weyant:2011hs}
A.~Weyant, M.~Wood-Vasey, L.~Wasserman, and P.~Freeman, {\it {An Unbiased
  Method of Modeling the Local Peculiar Velocity Field with Type Ia
  Supernovae}},  {\em Astrophys. J.} {\bf 732} (2011) 65,
  [\href{http://arxiv.org/abs/1103.1603}{{\tt arXiv:1103.1603}}].

\bibitem{Ma:2012tt}
Y.-Z. Ma and D.~Scott, {\it {Cosmic bulk flows on $50 {h}^{-1}$Mpc scales: A
  Bayesian hyper-parameter method and multi-shells likelihood analysis}},  {\em
  Mon. Not. Roy. Astron. Soc.} {\bf 428} (2013) 2017,
  [\href{http://arxiv.org/abs/1208.2028}{{\tt arXiv:1208.2028}}].

\bibitem{Rathaus:2013ut}
B.~Rathaus, E.~D. Kovetz, and N.~Itzhaki, {\it {Studying the Peculiar Velocity
  Bulk Flow in a Sparse Survey of Type-Ia SNe}},  {\em Mon. Not. Roy. Astron.
  Soc.} {\bf 431} (2013) 3678, [\href{http://arxiv.org/abs/1301.7710}{{\tt
  arXiv:1301.7710}}].

\bibitem{Feindt:2013pma}
U.~Feindt et~al., {\it {Measuring cosmic bulk flows with Type Ia Supernovae
  from the Nearby Supernova Factory}},  {\em Astron. Astrophys.} {\bf 560}
  (2013) A90, [\href{http://arxiv.org/abs/1310.4184}{{\tt arXiv:1310.4184}}].

\bibitem{Ma:2013oja}
Y.-Z. Ma and J.~Pan, {\it {An estimation of local bulk flow with the
  maximum-likelihood method}},  {\em Mon. Not. Roy. Astron. Soc.} {\bf 437}
  (2014), no.~2 1996--2004, [\href{http://arxiv.org/abs/1311.6888}{{\tt
  arXiv:1311.6888}}].

\bibitem{Johnson:2014kaa}
A.~Johnson et~al., {\it {The 6dF Galaxy Velocity Survey: Cosmological
  constraints from the velocity power spectrum}},  {\em Mon. Not. Roy. Astron.
  Soc.} {\bf 444} (2014) 3926, [\href{http://arxiv.org/abs/1404.3799}{{\tt
  arXiv:1404.3799}}].

\bibitem{Abate:2008au}
A.~Abate and O.~Lahav, {\it {The Three Faces of $\Omega_m$: Testing Gravity
  with Low and High Redshift SN Ia Surveys}},  {\em Mon. Not. Roy. Astron.
  Soc.} {\bf 389} (2008) 47, [\href{http://arxiv.org/abs/0805.3160}{{\tt
  arXiv:0805.3160}}].

\bibitem{Turnbull:2011ty}
S.~J. Turnbull, M.~J. Hudson, H.~A. Feldman, M.~Hicken, R.~P. Kirshner, and
  R.~Watkins, {\it {Cosmic flows in the nearby universe from Type Ia
  Supernovae}},  {\em Mon. Not. Roy. Astron. Soc.} {\bf 420} (2012) 447--454,
  [\href{http://arxiv.org/abs/1111.0631}{{\tt arXiv:1111.0631}}].

\bibitem{Castro:2014oja}
T.~{Castro} and M.~{Quartin}, {\it {First measurement of {$\sigma$}$_{8}$ using
  supernova magnitudes only}},  {\em Mon. Not. Roy. Astron. Soc.} {\bf 443}
  (2014) L6--L10, [\href{http://arxiv.org/abs/1403.0293}{{\tt
  arXiv:1403.0293}}].

\bibitem{Carrick:2015xza}
J.~Carrick, S.~J. Turnbull, G.~Lavaux, and M.~J. Hudson, {\it {Cosmological
  parameters from the comparison of peculiar velocities with predictions from
  the 2M++ density field}},  {\em Mon. Not. Roy. Astron. Soc.} {\bf 450}
  (2015), no.~1 317--332, [\href{http://arxiv.org/abs/1504.04627}{{\tt
  arXiv:1504.04627}}].

\bibitem{Schwarz:2007wf}
D.~J. Schwarz and B.~Weinhorst, {\it {(An)isotropy of the Hubble diagram:
  Comparing hemispheres}},  {\em Astron. Astrophys.} {\bf 474} (2007) 717--729,
  [\href{http://arxiv.org/abs/0706.0165}{{\tt arXiv:0706.0165}}].

\bibitem{Kalus:2012zu}
B.~Kalus, D.~J. Schwarz, M.~Seikel, and A.~Wiegand, {\it {Constraints on
  anisotropic cosmic expansion from supernovae}},  {\em Astron. Astrophys.}
  {\bf 553} (2013) A56, [\href{http://arxiv.org/abs/1212.3691}{{\tt
  arXiv:1212.3691}}].

\bibitem{Yang:2013gea}
X.~Yang, F.~Y. Wang, and Z.~Chu, {\it {Searching for a preferred direction with
  Union2.1 data}},  {\em Mon. Not. Roy. Astron. Soc.} {\bf 437} (2014), no.~2
  1840--1846, [\href{http://arxiv.org/abs/1310.5211}{{\tt arXiv:1310.5211}}].

\bibitem{Appleby:2014kea}
S.~Appleby, A.~Shafieloo, and A.~Johnson, {\it {Probing bulk flow with nearby
  SNe Ia data}},  {\em Astrophys. J.} {\bf 801} (2015), no.~2 76,
  [\href{http://arxiv.org/abs/1410.5562}{{\tt arXiv:1410.5562}}].

\bibitem{Lin:2015rza}
H.-N. Lin, S.~Wang, Z.~Chang, and X.~Li, {\it {Testing the isotropy of the
  Universe by using the JLA compilation of type-Ia supernovae}},  {\em Mon.
  Not. Roy. Astron. Soc.} {\bf 456} (2016), no.~2 1881--1885,
  [\href{http://arxiv.org/abs/1504.03428}{{\tt arXiv:1504.03428}}].

\bibitem{Javanmardi:2015sfa}
B.~Javanmardi, C.~Porciani, P.~Kroupa, and J.~Pflamm-Altenburg, {\it {Probing
  the isotropy of cosmic acceleration traced by Type Ia supernovae}},  {\em
  Astrophys. J.} {\bf 810} (2015) 47,
  [\href{http://arxiv.org/abs/1507.07560}{{\tt arXiv:1507.07560}}].

\bibitem{Bengaly:2015dza}
C.~A.~P. Bengaly, A.~Bernui, and J.~S. Alcaniz, {\it {Probing Cosmological
  Isotropy With Type IA Supernovae}},  {\em Astrophys. J.} {\bf 808} (2015) 39,
  [\href{http://arxiv.org/abs/1503.01413}{{\tt arXiv:1503.01413}}].

\bibitem{Hui:2005nm}
L.~Hui and P.~B. Greene, {\it {Correlated Fluctuations in Luminosity Distance
  and the (Surprising) Importance of Peculiar Motion in Supernova Surveys}},
  {\em Phys. Rev.} {\bf D73} (2006) 123526,
  [\href{http://arxiv.org/abs/astro-ph/0512159}{{\tt astro-ph/0512159}}].

\bibitem{Cooray:2006ft}
A.~Cooray and R.~R. Caldwell, {\it {Large-scale bulk motions complicate the
  hubble diagram}},  {\em Phys. Rev.} {\bf D73} (2006) 103002,
  [\href{http://arxiv.org/abs/astro-ph/0601377}{{\tt astro-ph/0601377}}].

\bibitem{Neill:2007fh}
{\bf SNLS} Collaboration, J.~D. Neill, M.~J. Hudson, and A.~J. Conley, {\it
  {The Peculiar Velocities of Local Type Ia Supernovae and their Impact on
  Cosmology}},  {\em Astrophys. J.} {\bf 661} (2007) L123,
  [\href{http://arxiv.org/abs/0704.1654}{{\tt arXiv:0704.1654}}].

\bibitem{Davis:2010jq}
T.~M. Davis et~al., {\it {The effect of peculiar velocities on supernova
  cosmology}},  {\em Astrophys. J.} {\bf 741} (2011) 67,
  [\href{http://arxiv.org/abs/1012.2912}{{\tt arXiv:1012.2912}}].

\bibitem{Cooray:2005yp}
A.~Cooray, D.~Holz, and D.~Huterer, {\it {Cosmology from supernova
  magnification maps}},  {\em Astrophys. J.} {\bf 637} (2006) L77--L80,
  [\href{http://arxiv.org/abs/astro-ph/0509579}{{\tt astro-ph/0509579}}].

\bibitem{Hannestad:2007fb}
S.~Hannestad, T.~Haugboelle, and B.~Thomsen, {\it {Precision measurements of
  large scale structure with future type Ia supernova surveys}},  {\em JCAP}
  {\bf 0802} (2008) 022, [\href{http://arxiv.org/abs/0705.0979}{{\tt
  arXiv:0705.0979}}].

\bibitem{Bhattacharya:2010cf}
S.~Bhattacharya, A.~Kosowsky, J.~A. Newman, and A.~R. Zentner, {\it {Galaxy
  Peculiar Velocities From Large-Scale Supernova Surveys as a Dark Energy
  Probe}},  {\em Phys. Rev.} {\bf D83} (2011) 043004,
  [\href{http://arxiv.org/abs/1008.2560}{{\tt arXiv:1008.2560}}].

\bibitem{Huterer:2015gpa}
D.~Huterer, D.~L. Shafer, and F.~Schmidt, {\it {No evidence for bulk velocity
  from type Ia supernovae}},  {\em JCAP} {\bf 1512} (2015), no.~12 033,
  [\href{http://arxiv.org/abs/1509.04708}{{\tt arXiv:1509.04708}}].

\bibitem{S14b}
D.~Scolnic et~al., {\it {Systematic Uncertainties Associated with the
  Cosmological Analysis of the First Pan-STARRS1 Type Ia Supernova Sample}},
  {\em Astrophys. J.} {\bf 795} (2014), no.~1 45,
  [\href{http://arxiv.org/abs/1310.3824}{{\tt arXiv:1310.3824}}].

\bibitem{Supercal}
D.~Scolnic et~al., {\it {SUPERCAL: Cross=Calibration of Multiple Photometric
  Systems to Improve Cosmological Measurements with type Ia Supernovae}},  {\em
  Astrophys. J.} {\bf 815} (2015), no.~2 117.

\bibitem{ScolnicKessler16}
D.~Scolnic and R.~Kessler, {\it {Measuring Type Ia Supernova Populations of
  Stretch and Color and Predicting Distance Biases}},  {\em Astrophys. J.} {\bf
  822} (2016), no.~2 L35, [\href{http://arxiv.org/abs/1603.01559}{{\tt
  arXiv:1603.01559}}].

\bibitem{Stritzinger11}
M.~D. Stritzinger et~al., {\it {The Carnegie Supernova Project: Second
  Photometry Data Release of Low-Redshift Type Ia Supernovae}},  {\em Astron.
  J.} {\bf 142} (2011) 156, [\href{http://arxiv.org/abs/1108.3108}{{\tt
  arXiv:1108.3108}}].

\bibitem{Hicken12}
M.~Hicken et~al., {\it {CfA4: Light Curves for 94 Type Ia Supernovae}},  {\em
  Astrophys. J. Suppl.} {\bf 200} (2012) 12,
  [\href{http://arxiv.org/abs/1205.4493}{{\tt arXiv:1205.4493}}].

\bibitem{Guy:2007dv}
{\bf SNLS} Collaboration, J.~Guy et~al., {\it {SALT2: Using distant supernovae
  to improve the use of Type Ia supernovae as distance indicators}},  {\em
  Astron. Astrophys.} {\bf 466} (2007) 11--21,
  [\href{http://arxiv.org/abs/astro-ph/0701828}{{\tt astro-ph/0701828}}].

\bibitem{Betoule:2014frx}
{\bf SDSS} Collaboration, M.~Betoule et~al., {\it {Improved cosmological
  constraints from a joint analysis of the SDSS-II and SNLS supernova
  samples}},  {\em Astron. Astrophys.} {\bf 568} (2014) A22,
  [\href{http://arxiv.org/abs/1401.4064}{{\tt arXiv:1401.4064}}].

\bibitem{Kessler:2009ys}
R.~Kessler et~al., {\it {First-year Sloan Digital Sky Survey-II (SDSS-II)
  Supernova Results: Hubble Diagram and Cosmological Parameters}},  {\em
  Astrophys. J. Suppl.} {\bf 185} (2009) 32--84,
  [\href{http://arxiv.org/abs/0908.4274}{{\tt arXiv:0908.4274}}].

\bibitem{Dressler:1987ny}
A.~Dressler, D.~Lynden-Bell, D.~Burstein, R.~L. Davies, S.~M. Faber,
  R.~Terlevich, and G.~Wegner, {\it {Spectroscopy and photometry of elliptical
  galaxies. 1. A New distance estimator}},  {\em Astrophys. J.} {\bf 313}
  (1987) 42--58.

\bibitem{Djorgovski:1987vx}
S.~Djorgovski and M.~Davis, {\it {Fundamental properties of elliptical
  galaxies}},  {\em Astrophys. J.} {\bf 313} (1987) 59.

\bibitem{Jones:2004zy}
D.~H. Jones et~al., {\it {The 6dF Galaxy Survey: Samples, observational
  techniques and the first data release}},  {\em Mon. Not. Roy. Astron. Soc.}
  {\bf 355} (2004) 747--763, [\href{http://arxiv.org/abs/astro-ph/0403501}{{\tt
  astro-ph/0403501}}].

\bibitem{Jones:2009yz}
D.~H. Jones et~al., {\it {The 6dF Galaxy Survey: Final Redshift Release (DR3)
  and Southern Large-Scale Structures}},  {\em Mon. Not. Roy. Astron. Soc.}
  {\bf 399} (2009) 683, [\href{http://arxiv.org/abs/0903.5451}{{\tt
  arXiv:0903.5451}}].

\bibitem{Beutler:2011hx}
F.~Beutler, C.~Blake, M.~Colless, D.~H. Jones, L.~Staveley-Smith, L.~Campbell,
  Q.~Parker, W.~Saunders, and F.~Watson, {\it {The 6dF Galaxy Survey: Baryon
  Acoustic Oscillations and the Local Hubble Constant}},  {\em Mon. Not. Roy.
  Astron. Soc.} {\bf 416} (2011) 3017--3032,
  [\href{http://arxiv.org/abs/1106.3366}{{\tt arXiv:1106.3366}}].

\bibitem{Magoulas:2012jy}
C.~Magoulas, C.~M. Springob, M.~Colless, D.~H. Jones, L.~A. Campbell, J.~R.
  Lucey, J.~Mould, T.~Jarrett, A.~Merson, and S.~Brough, {\it {The 6dF Galaxy
  Survey: The Near-Infrared Fundamental Plane of Early-Type Galaxies}},  {\em
  Mon. Not. Roy. Astron. Soc.} {\bf 427} (2012) 245,
  [\href{http://arxiv.org/abs/1206.0385}{{\tt arXiv:1206.0385}}].

\bibitem{Springob:2014qja}
C.~M. Springob, C.~Magoulas, M.~Colless, J.~Mould, P.~Erdogdu, D.~H. Jones,
  J.~R. Lucey, L.~Campbell, and C.~J. Fluke, {\it {The 6dF Galaxy Survey:
  Peculiar Velocity Field and Cosmography}},  {\em Mon. Not. Roy. Astron. Soc.}
  {\bf 445} (2014), no.~3 2677--2697,
  [\href{http://arxiv.org/abs/1409.6161}{{\tt arXiv:1409.6161}}].

\bibitem{Campbell:2014uia}
L.~A. Campbell et~al., {\it {The 6dF Galaxy Survey: Fundamental Plane Data}},
  {\em Mon. Not. Roy. Astron. Soc.} {\bf 443} (2014), no.~2 1231--1251,
  [\href{http://arxiv.org/abs/1406.4867}{{\tt arXiv:1406.4867}}].

\bibitem{Lewis:1999bs}
A.~Lewis, A.~Challinor, and A.~Lasenby, {\it {Efficient computation of CMB
  anisotropies in closed FRW models}},  {\em Astrophys. J.} {\bf 538} (2000)
  473--476, [\href{http://arxiv.org/abs/astro-ph/9911177}{{\tt
  astro-ph/9911177}}].

\bibitem{Smith:2002dz}
{\bf VIRGO Consortium} Collaboration, R.~E. Smith, J.~A. Peacock, A.~Jenkins,
  S.~D.~M. White, C.~S. Frenk, F.~R. Pearce, P.~A. Thomas, G.~Efstathiou, and
  H.~M.~P. Couchmann, {\it {Stable clustering, the halo model and nonlinear
  cosmological power spectra}},  {\em Mon. Not. Roy. Astron. Soc.} {\bf 341}
  (2003) 1311, [\href{http://arxiv.org/abs/astro-ph/0207664}{{\tt
  astro-ph/0207664}}].

\bibitem{Takahashi:2012em}
R.~Takahashi, M.~Sato, T.~Nishimichi, A.~Taruya, and M.~Oguri, {\it {Revising
  the Halofit Model for the Nonlinear Matter Power Spectrum}},  {\em Astrophys.
  J.} {\bf 761} (2012) 152, [\href{http://arxiv.org/abs/1208.2701}{{\tt
  arXiv:1208.2701}}].

\bibitem{Ade:2015xua}
{\bf Planck} Collaboration, P.~A.~R. Ade et~al., {\it {Planck 2015 results.
  XIII. Cosmological parameters}},  {\em Astron. Astrophys.} {\bf 594} (2016)
  A13, [\href{http://arxiv.org/abs/1502.01589}{{\tt arXiv:1502.01589}}].

\bibitem{Beutler:2012px}
F.~Beutler, C.~Blake, M.~Colless, D.~H. Jones, L.~Staveley-Smith, G.~B. Poole,
  L.~Campbell, Q.~Parker, W.~Saunders, and F.~Watson, {\it {The 6dF Galaxy
  Survey: $z \approx 0$ measurement of the growth rate and $\sigma_8$}},  {\em
  Mon. Not. Roy. Astron. Soc.} {\bf 423} (2012) 3430--3444,
  [\href{http://arxiv.org/abs/1204.4725}{{\tt arXiv:1204.4725}}].

\bibitem{Blake:2013nif}
C.~Blake et~al., {\it {Galaxy And Mass Assembly (GAMA): improved cosmic growth
  measurements using multiple tracers of large-scale structure}},  {\em Mon.
  Not. Roy. Astron. Soc.} {\bf 436} (2013) 3089,
  [\href{http://arxiv.org/abs/1309.5556}{{\tt arXiv:1309.5556}}].

\bibitem{Blake:2011rj}
C.~Blake et~al., {\it {The WiggleZ Dark Energy Survey: the growth rate of
  cosmic structure since redshift z=0.9}},  {\em Mon. Not. Roy. Astron. Soc.}
  {\bf 415} (2011) 2876, [\href{http://arxiv.org/abs/1104.2948}{{\tt
  arXiv:1104.2948}}].

\bibitem{Beutler:2016arn}
{\bf BOSS} Collaboration, F.~Beutler et~al., {\it {The clustering of galaxies
  in the completed SDSS-III Baryon Oscillation Spectroscopic Survey:
  Anisotropic galaxy clustering in Fourier-space}},  {\em Submitted to: Mon.
  Not. Roy. Astron. Soc.} (2016) [\href{http://arxiv.org/abs/1607.03150}{{\tt
  arXiv:1607.03150}}].

\bibitem{delaTorre:2013rpa}
S.~de~la Torre et~al., {\it {The VIMOS Public Extragalactic Redshift Survey
  (VIPERS). Galaxy clustering and redshift-space distortions at $z=0.8$ in the
  first data release}},  {\em Astron. Astrophys.} {\bf 557} (2013) A54,
  [\href{http://arxiv.org/abs/1303.2622}{{\tt arXiv:1303.2622}}].

\bibitem{Hyde:2008yh}
J.~B. Hyde and M.~Bernardi, {\it {The luminosity and stellar mass Fundamental
  Plane of early-type galaxies}},  {\em Mon. Not. Roy. Astron. Soc.} {\bf 396}
  (2009) 1171, [\href{http://arxiv.org/abs/0810.4924}{{\tt arXiv:0810.4924}}].

\bibitem{LaBarbera:2009vn}
F.~La~Barbera, R.~R. de~Carvalho, I.~G. de~la Rosa, and P.~A.~A. Lopes, {\it
  {SPIDER II - The Fundamental Plane of Early-type Galaxies in grizYJHK}},
  {\em Mon. Not. Roy. Astron. Soc.} {\bf 408} (2010) 1335,
  [\href{http://arxiv.org/abs/0912.4558}{{\tt arXiv:0912.4558}}].

\bibitem{Robotham:2015fia}
A.~S.~G. Robotham and D.~Obreschkow, {\it {Hyper-Fit: Fitting Linear Models to
  Multidimensional Data with Multivariate Gaussian Uncertainties}},  {\em Publ.
  Astron. Soc. Austral.} {\bf 32} (2015) 33,
  [\href{http://arxiv.org/abs/1508.02145}{{\tt arXiv:1508.02145}}].

\bibitem{Amanullah:2010vv}
R.~Amanullah et~al., {\it {Spectra and Light Curves of Six Type Ia Supernovae
  at $0.511 < z < 1.12$ and the Union2 Compilation}},  {\em Astrophys. J.} {\bf
  716} (2010) 712--738, [\href{http://arxiv.org/abs/1004.1711}{{\tt
  arXiv:1004.1711}}].

\bibitem{1993ApJ...405..437N}
A.~{Nusser} and A.~{Dekel}, {\it {Omega and the initial fluctuations from
  velocity and density fields}},  {\em Astrophys. J.} {\bf 405} (Mar., 1993)
  437--448.

\bibitem{Bernardeau:1994vz}
F.~Bernardeau, R.~Juszkiewicz, A.~Dekel, and F.~R. Bouchet, {\it {Omega from
  the skewness of the cosmic velocity divergence}},  {\em Mon. Not. Roy.
  Astron. Soc.} {\bf 274} (1995) 20--26,
  [\href{http://arxiv.org/abs/astro-ph/9404052}{{\tt astro-ph/9404052}}].

\bibitem{Dekel:1993hq}
A.~Dekel and M.~J. Rees, {\it {Omega from velocities in voids}},  {\em
  Astrophys. J.} {\bf 422} (1994) L1,
  [\href{http://arxiv.org/abs/astro-ph/9308029}{{\tt astro-ph/9308029}}].

\bibitem{2005ARNPS..55..141K}
J.~R. {Klein} and A.~{Roodman}, {\it {Blind Analysis in Nuclear and Particle
  Physics}},  {\em Annual Review of Nuclear and Particle Science} {\bf 55}
  (Dec., 2005) 141--163.

\bibitem{Johnston:2007yh}
{\bf ASKAP} Collaboration, S.~Johnston et~al., {\it {Science With The
  Australian Square Kilometre Array Pathfinder}},  {\em PoS} {\bf MRU} (2007)
  006, [\href{http://arxiv.org/abs/0711.2103}{{\tt arXiv:0711.2103}}]. [Publ.
  Astron. Soc. Austral.24,174(2007)].

\bibitem{Koribalski:2012pf}
B.~S. Koribalski, {\it {Overview on spectral line source finding and
  visualisation}},  {\em Publ. Astron. Soc. Austral.} {\bf 29} (2012) 359,
  [\href{http://arxiv.org/abs/1206.6916}{{\tt arXiv:1206.6916}}].

\bibitem{Duffy:2012gr}
A.~R. Duffy, M.~J. Meyer, L.~Staveley-Smith, M.~Bernyk, D.~J. Croton, B.~S.
  Koribalski, D.~Gerstmann, and S.~Westerlund, {\it {Predictions for ASKAP
  Neutral Hydrogen Surveys}},  {\em Mon. Not. Roy. Astron. Soc.} {\bf 426}
  (2012) 3385, [\href{http://arxiv.org/abs/1208.5592}{{\tt arXiv:1208.5592}}].

\bibitem{Koda:2013eya}
J.~Koda, C.~Blake, T.~Davis, C.~Magoulas, C.~M. Springob, M.~Scrimgeour,
  A.~Johnson, G.~B. Poole, and L.~Staveley-Smith, {\it {Are peculiar velocity
  surveys competitive as a cosmological probe?}},  {\em Mon. Not. Roy. Astron.
  Soc.} {\bf 445} (2014), no.~4 4267--4286,
  [\href{http://arxiv.org/abs/1312.1022}{{\tt arXiv:1312.1022}}].

\bibitem{Howlett:2016urc}
C.~Howlett, L.~Staveley-Smith, and C.~Blake, {\it {Cosmological Forecasts for
  Combined and Next Generation Peculiar Velocity Surveys}},  {\em Mon. Not.
  Roy. Astron. Soc.} {\bf 464} (2017), no.~3 2517--2544--2544,
  [\href{http://arxiv.org/abs/1609.08247}{{\tt arXiv:1609.08247}}].

\end{thebibliography}\endgroup

\end{document}